\documentclass[12pt]{article}
\usepackage{amsmath}
\usepackage{amssymb}
\usepackage{bm}
\newcommand{\be}{\begin{equation}}
\newcommand{\ee}{\end{equation}}
\newcommand{\bea}{\begin{eqnarray}}
\newcommand{\eea}{\end{eqnarray}}

\newcommand{\mbts}[1]{_{\mbox{\tiny #1}}}
\newcommand{\mbsu}[1]{\mbox{\scriptsize #1}}

\newcommand{\ds}{\displaystyle}
\newcommand{\scs}{\scriptstyle}
\newcommand{\txts}{\textstyle}
\newcommand{\ve}{\varepsilon}

\newcommand{\vphi}{\varphi}
\newcommand{\mbold}[1]{\mbox{\boldmath $#1$}}
\newcommand{\bnabla}{\mbold{\nabla}}
\numberwithin{equation}{section}
\makeatletter
\renewcommand{\section}{\@startsection{section}{1}{0pt}%
{-3.5ex plus -1ex minus -.2ex}{2.3ex plus.2ex}%
{\normalsize\bf}}
\renewcommand{\subsection}{\@startsection{subsection}{1}{0pt}%
{-3.5ex plus -1ex minus -.2ex}{2.3ex plus.2ex}%
{\normalsize\bf}}
\makeatother
\begin{document}
\title{%
\large\bf
 QUASIPARTICLE TIME BLOCKING APPROXIMATION\\
 WITHIN THE FRAMEWORK OF GENERALIZED\\
 GREEN FUNCTION FORMALISM}
\author{%
V. I. Tselyaev \\
\vspace{-0.5em}
{\it \normalsize
 Nuclear Physics Department,
 V. A. Fock Institute of Physics,}\\
{\it \normalsize
 St. Petersburg State University, 198504,
 St. Petersburg, Russia}}
\date{May 11, 2005}
\maketitle
\begin{abstract}
The problem of the microscopic description of excited states
of the even-even open-shell atomic nuclei
is considered. A model is formulated which allows one to go
beyond the quasiparticle random phase approximation.
The physical content of the model is determined by the
quasiparticle time blocking approximation (QTBA) which
enables one to include contributions of the two-quasiparticle
and the two-phonon configurations, while excluding (blocking)
more complicated intermediate states. In addition, the QTBA
ensures consistent treatment of ground state correlations
in the Fermi systems with pairing.
The model is based on the generalized Green function formalism
(GGFF) in which the normal and the anomalous Green functions
are treated in a unified way in terms of the components of
generalized Green functions in a doubled space.
Modification of the GGFF is considered
in the case when the many-body nuclear Hamiltonian
contains two-, three-, and other many-particle effective forces.
\vspace{2em}
\begin{flushleft}
\end{flushleft}
\end{abstract}
\newpage
\section{INTRODUCTION}

One of the most widely used approaches which are applied to
the description of excitations of the even-even atomic nuclei
is the random phase approximation
(RPA, see, e.g., Ref.~\cite{RS}). Within this approximation
the nuclear excitations are treated as the one-phonon states
which are superpositions of the one-particle-one-hole
(1p1h) configurations. However, this approach
is applicable only in the case when pairing correlations are
not essential, i.e., strictly speaking, only for magic nuclei.
Generalization of the RPA taking into account pairing
correlations explicitly is the quasiparticle RPA (QRPA),
where the excited states (phonons) are expanded in the
two-quasiparticle (2q) configurations. Thereby the QRPA extends
the range of the RPA to the open-shell (non-magic) nuclei.
Nevertheless, despite the significant success of both
the RPA and the QRPA, there are several reasons for development
of the models going beyond these approximations.

First of all, description of the nuclear excitations in terms
of the one-phonon wave functions is justified only for low-lying
states. At higher excitation energies,
fragmentation of the one-phonon states becomes important.
This means that in addition to the 1p1h or 2q configurations
the more complex configurations should be incorporated
(see Ref.~\cite{S92}). The role of the effects related to the
complex configurations is well manifested, for example,
in the theory of giant multipole resonances (GMRs).
It is well known (see, e.g., Ref.~\cite{KTT})
that RPA and QRPA enable one to describe the centroid energies
and total strengths of the GMRs.
However, both models fail to reproduce the
total widths of the resonances and their fine structure.
The reason is that these characteristics of the GMRs are
significantly affected by the complex (mainly 2p2h or 4q)
configurations which form the spreading width of the
resonance.

Another direction of developing a nuclear structure theory
is associated with the models
in which ground state correlations (GSC) beyond
the RPA and QRPA GSC are taken into account
(see Refs.~\cite{KTT,TSA,DNSW,KVC}). It has been shown that
the GSC caused by complex configurations play
an important role in the theoretical description of the
experimental data. In what follows we will refer to this
type of GSC as the GSC2 in order to distinguish them from
the GSC1 included in the RPA and QRPA.
In some cases, the GSC2 can strongly affect
the transition strengths and can even lead to the appearance
of new transitions which are absent in the calculations
including complex configurations in the excited states
only, i.e. in the calculations neglecting this type of GSC
and using a restricted basis (see Ref.~\cite{KTT}).

A variety of models have been developed to study the effects
of complex configurations on the structure of excited states
of the even-even atomic nuclei
(in addition to the aforementioned papers
see also Refs. \cite{BBBD,BB,AY,T89,CBGBB,CB01,SBC}
and references therein).
Nevertheless, until recently (see Refs. \cite{CB01,SBC}),
the quasiparticle-phonon model (QPM, Ref.~\cite{S92})
developed by Soloviev and co-workers was the only working
approach which consistently treats the complex configurations
and the pairing correlations on an equal footing.
It is no surprising that comprehensive studies of the
excitations of the open-shell nuclei taking into account
complex (mainly two-phonon) configurations at the microscopic
level have been carried out only within the QPM.
In view of this the development of other approaches
in this direction is particularly important.

The principal goal of the present paper is to generalize
the model of Ref.~\cite{T89} by including the pairing
correlations. This model was developed to describe the excited
states of the even-even doubly magic nuclei taking into
account 2p2h (more precisely, 1p1h$\otimes$phonon)
configurations. The model is based on the Green function (GF)
formalism. The GSC2 were completely included within the
model approximations. In the framework of this model the
calculations of the GMRs in magic stable and unstable nuclei
have been performed. Some of the results are presented in
Refs. \cite{KTT,T89}. In these calculations reasonable
agreement with the experimental data for the integral
characteristics of GMRs, including the total resonance widths,
has been obtained. Thus, one can expect that the extension of
the model \cite{T89} to the open-shell nuclei
will also give reasonable results.

The second goal of the paper is to provide a modification
of the GF formalism for the Fermi systems with pairing
which is most suitable for solving the problem under
consideration. To this aim,
the generalized Green function formalism (GGFF)
is presented, in which the normal and the anomalous GFs
are treated in a unified way in terms of the components
of generalized GFs in a doubled space.
Modification of the GGFF is considered in the case
when the many-body nuclear Hamiltonian contains two-, three-,
and other many-particle effective forces.

The paper is divided into two main parts.
The first part (Sec.~\ref{sec2}) reviews
the basic formulas and equations of the GGFF.
The single-quasiparticle basis functions,
which provide suitable representation of
the model equations, are introduced.
The second part (Sec.~\ref{sec3}) contains the
formulation of the model in which
pairing correlations, 2q, 2q$\otimes$phonon, and
two-phonon configurations are included.
The model is analyzed within sum rule approach.
The conclusions are given in the last section.

\section{GENERALIZED GREEN FUNCTION FORMALISM \label{sec2}}

\subsection{Basic definitions \label{sec21}}

Let $a^{\dag} (x)$ and $a (x)$ be creation and annihilation
operators of particles (free fermions) in the
coordinate representation of the usual single-particle space.
Here symbol $x = \{{\mbold r}, \sigma, \tau \}$
includes the spatial coordinate $\mbold r$, the
spin $\sigma$, and the isospin $\tau$ variables.
Considering the Fermi systems with pairing correlations
it is convenient
to pass from this single-particle space spanned by the
coordinates $x$ to the extended (doubled) space spanned by
the coordinates $y = \{ x, \chi \}$, where $\chi = \pm 1$
is an additional index introduced for denoting the different
components of the single-particle functions in the extended
space (see Refs.~\cite{RS,KSTV} for details).
Let us define the operators $b (y) = b (x,\chi)$
by the relations
\be
b (x,+) = a (x)\,,\qquad b (x,-) = a^{\dag} (x)\,.
\label{dfb}
\ee
From this it follows that $b^{\dag} (y) = b (\bar{y})$
where $\bar{y} = \{ x, -\chi \}$.
The Heisenberg representation of the $b$-operators
(in units where Planck's constant $\hbar =1$) reads:
\be
\Psi (z) = e^{i H t}\, b (y)\, e^{- i H t}\,.
\label{bhr}
\ee
Here and in the following $z = \{ t, y \}$,
$\,t$ is the time variable, $H$ is a many-body Hamiltonian
of an interacting fermion system.
Obviously, these $\Psi$-operators possess the property:
\be
\Psi^{\dag} (z) = \Psi (\bar{z})\,,
\label{prpsi}
\ee
where $\bar{z} = \{ t, \bar{y} \}$.

We will assume that the motion in the fermion system
is determined by the nonrelativistic Hamiltonian $H$ of the form
\be
H=H^0 + V\,,
\label{dfhtot}
\ee
where $H^0$ is a single-particle Hamiltonian including
the external anomalous pair potentials:
\bea
H^0 &=& \int dx^{\vphantom{'}}_1\, dx'_1\, \biggl(\,
h^0(x^{\vphantom{'}}_1,x'_1)\,
a^{\dag}(x^{\vphantom{'}}_1)\, a(x'_1)
\nonumber\\
&&+\;
{\txts \frac{1}{2}}\,\Delta^0(x^{\vphantom{'}}_1,x'_1)\,
a^{\dag}(x^{\vphantom{'}}_1)\,a^{\dag}(x'_1)
- {\txts \frac{1}{2}}\,
\Delta^{0^{\ds *}}(x^{\vphantom{'}}_1,x'_1)\,
a(x^{\vphantom{'}}_1)\,a(x'_1)\,\biggr)\,,
\label{dfh0}
\eea
$V$ is an interaction including two-, three-, and other
many-particle effective forces:
\be
V=\sum_{k=2}^K V^{\,(k)}\,,
\label{dfv}
\ee
\bea
V^{\,(k)} &=& \frac{1}{k\,!}\,
\int dx^{\vphantom{'}}_1 \cdot \ldots \cdot dx^{\vphantom{'}}_k\,
dx'_1 \cdot \ldots \cdot dx'_k\;
v^{\,(k)}(x^{\vphantom{'}}_1,\ldots , x^{\vphantom{'}}_k\,;\,
x'_1,\ldots , x'_k)
\nonumber\\
&&\times\;
a^{\dag} (x^{\vphantom{'}}_1)\cdot \ldots \cdot
a^{\dag} (x^{\vphantom{'}}_k)\,
a (x'_k)\cdot \ldots \cdot a (x'_1)\,.
\label{dfvk}
\eea
Here and in the following
$\int dx$ means the space integral over ${\mbold r}$
and the sum over $\sigma$ and $\tau$ indices. Analogously, in the
following $\int dy$ will denote $\int dx$ and the sum over $\chi$,
$\int dz$ will denote $\int dt\,dy$.
In case of the exact nuclear Hamiltonian we have:
\be
h^0(x,x') = - \left( \frac{\bnabla^2_{\bf r}}{2m}
+ \mu_{\tau} \right) \delta(x,x')\,,
\qquad \Delta^0(x,x') = 0\,,
\label{dfhd0}
\ee
where
$\delta (x,x') = \delta({\mbold r}-{\mbold r}') \,
\delta_{\sigma,\sigma'}\, \delta_{\tau,\tau'}$, $\;\mu_{\tau}$
is the chemical potential for the nucleons with the isospin
projection $\tau$
which is introduced to simplify the following equations.
Notice that the Hamiltonian $H^0$ can be formally
rewritten in terms of the $b$-operators as
\be
H^0 = \frac{1}{2}\,\int dy\, dy'\; {\cal H}^0(y,y')\;
b^{\dag}(y)\, b(y')\, + \,\epsilon_0\,,
\label{dfh0b}
\ee
where
\be
\left.
\begin{array}{ll}
{\cal H}^0(x,+\, ,\, x',+) = h^0 (x,x')\,, &
{\cal H}^0(x,+\, ,\, x',-) = \Delta^0 (x,x')\,, \\
{\cal H}^0(x,-\, ,\, x',+) = - \Delta^{0^{\ds *}} (x,x')\,,\quad &
{\cal H}^0(x,-\, ,\, x',-) = - h^{0^{\ds *}} (x,x')\,, \\
\end{array}
\right\}
\label{dfh0m}
\ee
\be
\epsilon_0 = \frac{1}{2}\,\int dx\, h^0 (x,x)\,.
\label{dfeps}
\ee

Let $|\, 0 \,\rangle$ be the wave function of the ground state
of the interacting fermion system. If the Hamiltonian $H$
does not contain the external anomalous pair potentials
$\Delta^0$ [e.g., if Eqs.~(\ref{dfhd0}) are fulfilled],
the number of particles is conserved, and $|\, 0 \,\rangle$
is an eigenfunction of the particle-number operator.
However, in general case we shall not suppose that
$\Delta^0 = 0$ in Eq.~(\ref{dfh0}),
i.e. we shall not suppose that the condition of
the particle-number conservation is fulfilled for
$|\, 0 \,\rangle$. Let us define the $k$-particle generalized
Green function (GF) in the time representation by the formula:
\be
G^{\,(k)}(z^{\vphantom{'}}_1,\ldots , z^{\vphantom{'}}_k\,;\,
z'_1,\ldots , z'_k) =
i^{-k}\,\langle\, 0\, |\, \mbox{T}\,
\Psi (z^{\vphantom{'}}_1)\cdot \ldots \cdot
\Psi (z^{\vphantom{'}}_k)\,
\Psi^{\dag} (z'_k)\cdot \ldots \cdot \Psi^{\dag} (z'_1)\,
|\, 0 \,\rangle \,,
\label{dfgfk}
\ee
where $\mbox{T}$ is the time-ordering operator. In particular,
for the single-particle GF we have:
\be
G(z, z') \equiv G^{\,(1)}(z\,;\, z') =
-i\,\langle\, 0\, |\, \mbox{T}\,
\Psi (z)\, \Psi^{\dag} (z')\,|\, 0 \,\rangle \,.
\label{dfgf1}
\ee
The property
\be
G(z, z') = - G(\bar{z}', \bar{z})
\label{prgf1}
\ee
follows from Eqs.~(\ref{prpsi}) and (\ref{dfgf1}).
It can be seen from the definitions (\ref{dfb}) and (\ref{bhr})
that the normal and the anomalous GFs are the components
of the generalized GFs $G^{\,(k)}$ corresponding to the
different values of the $\chi$-indices.

\subsection{Equations of motion for the Green functions
            \label{sec22}}

In case of the Fermi systems with pairing,
the equations of motion for the many-particle GFs can be obtained
with the help of the same technique based on the
generating functionals depending on the auxiliary source fields
which is frequently used for the Fermi systems without pairing
correlations (see, e.g., Ref~\cite{SWW}).
Let us define the generating functional $W$ depending
on the source field $\xi$ as
\be
W[\,\xi\,] = \ln \langle\, 0\, |\, \mbox{T}\,\mbox{U}
\,|\, 0 \,\rangle \,,
\label{dfgfw}
\ee
where
\be
\mbox{U} = \exp \left( i \int dz\, dz'\; \xi (z,z')\;
\Psi^{\dag}(z)\, \Psi(z') \right)\,.
\label{dfu}
\ee
It follows from Eq.~(\ref{prpsi}) that one can consider
the equality
\be
\xi (z,z') = - \xi (\bar{z}',\bar{z})
\label{prpxi}
\ee
to be fulfilled.
Let us introduce the GFs with a source field $\xi$:
\be
G^{\,(k)}_{\xi}(z^{\vphantom{'}}_1,\ldots ,
z^{\vphantom{'}}_k\,;\, z'_1,\ldots , z'_k) =
i^{-k}\,\frac{\langle\, 0\, |\, \mbox{T}\,\mbox{U}
\Psi (z^{\vphantom{'}}_1)\cdot \ldots \cdot
\Psi (z^{\vphantom{'}}_k)\,
\Psi^{\dag} (z'_k)\cdot \ldots \cdot \Psi^{\dag} (z'_1)\,
|\, 0 \,\rangle}
{\langle\, 0\, |\, \mbox{T}\,\mbox{U}\,|\, 0 \,\rangle} \,.
\label{dfgfkx}
\ee
In particular, we have:
\be
G_{\xi}(z, z') \equiv G^{\,(1)}_{\xi}(z\,;\, z') =
-i\,\frac{\langle\, 0\, |\, \mbox{T}\,\mbox{U}
\Psi (z)\, \Psi^{\dag} (z')\,|\, 0 \,\rangle}
{\langle\, 0\, |\, \mbox{T}\,\mbox{U}\,|\, 0 \,\rangle}
= - G_{\xi}(\bar{z}', \bar{z})\,.
\label{dfgf1x}
\ee
Obviously, $G^{\,(k)}_{\xi}$ coincides with $G^{\,(k)}$ defined
by Eq.~(\ref{dfgfk}) at $\xi=0$.

It is easy to see that the GFs $G^{\,(k)}_{\xi}$ can be obtained
from the generating functional $W[\,\xi\,]$ by a successive
differentiation with respect to $\xi$. In particular, we obtain:
\be
G_{\xi}(z_1, z_2) =
\frac{\delta W}{\delta \xi (z_2, z_1)}\,,
\label{dxi1}
\ee
\be
L_{\xi}(z_1, z_2;\,z_3, z_4) =
\frac{\delta^2 W}{\delta \xi (z_1, z_2)\,
\delta \xi (z_4, z_3)} =
\frac{\delta G_{\xi}(z_2, z_1)}
{\delta \xi (z_4, z_3)}\,,
\label{dxi2}
\ee
where $L_{\xi}$ is the response function defined as
\be
L_{\xi}(z_1, z_2;\,z_3, z_4) =
G^{\,(2)}_{\xi}(z_2, z_3;\,z_1, z_4)-
G_{\xi}(z_2, z_1)\, G_{\xi}(z_3, z_4)
\label{dfrf}
\ee
(notice that this formula differs from the definition in
Ref~\cite{SWW} by the permutation of the arguments of $L_{\xi}$).

The equation of motion for the single-particle GF
is obtained by the differentiation of $G_{\xi}(z_1, z_2)$
with respect to time
in analogy to the case of the Fermi systems without pairing
correlations (see Ref~\cite{SWW}). It has the form
\be
\left( G^0 \right)^{-1}\! (z_1, z_2)
+ \xi (z_1, z_2) - \xi (\bar{z}_2, \bar{z}_1) =
G^{-1}_{\xi} (z_1, z_2) + \Sigma_{\,\xi} (z_1, z_2)\,,
\label{eqmgf1}
\ee
where
\be
\left( G^0 \right)^{-1}\! (z_1, z_2) =
\left( i\, \delta (y_1, y_2)\,
\frac{\partial}{\partial t_1}
- {\cal H}^0 (y_1, y_2) \right) \delta (t_1 - t_2)\,,
\label{dfgf0}
\ee
$\delta (y, y') = \delta_{\chi, \chi'}\, \delta (x, x')$,
$\;\Sigma_{\,\xi}$ is the mass operator which is defined
by the equations:
\bea
\int dz''\, \Sigma_{\,\xi} (z, z'')\,
G_{\xi} (z'', z') &=&
\sum_{k=2}^K \frac{i^{1-k}}{k\,!\,(k-1)}
\int dz'_2 \cdot \ldots \cdot dz'_k\,
dz''_1 \cdot \ldots \cdot dz''_k\,
\nonumber\\
&&\times\;
{\cal W}^{\,(k)}
(z,z'_2, \ldots , z'_k\,;\, z''_1,\ldots , z''_k)
\nonumber\\
&&\times\;
G^{\,(k)}_{\xi}
(z''_1,\ldots , z''_k\,;\,z',z'_2,\ldots , z'_k)\,,
\label{dfsigx}\\
{\cal W}^{\,(k)}
(z^{\vphantom{'}}_1, \ldots , z^{\vphantom{'}}_k\,;
\, z'_1,\ldots , z'_k) &=&
\delta_{\chi^{\vphantom{'}}_1,\,\chi'_1}\,
\delta (t^{\vphantom{'}}_1 - t'_1 - \chi^{\vphantom{'}}_1 \cdot 0)
\nonumber\\
&&\times\;
\left( \prod_{l=1}^{k-1}
\delta_{\chi^{\vphantom{'}}_l,\,\chi^{\vphantom{'}}_{l+1}}\,
\delta_{\chi'^{\vphantom{'}}_l,\,\chi'_{l+1}}\,
\delta (t^{\vphantom{'}}_l - t^{\vphantom{'}}_{l+1})\,
\delta (t'_l - t'_{l+1}) \right)
\nonumber\\
&&\times\;
\bigl[\, \delta_{\chi^{\vphantom{'}}_1,\,+1}\,
w^{\,(k)} (x^{\vphantom{'}}_1, \ldots , x^{\vphantom{'}}_k\,;
\, x'_1,\ldots , x'_k)
\nonumber\\
&&+\;\,
\delta_{\chi^{\vphantom{'}}_1,\,-1}\, (-1)^k\,
w^{\,(k)} (x'_1, \ldots , x'_k\,;\,
x^{\vphantom{'}}_1,\ldots , x^{\vphantom{'}}_k)\,\bigr]\,.
\label{dfwkz}
\eea
In the Eq.~(\ref{dfwkz}), $w^{\,(k)}$ is the antisymmetrized
matrix element of the $k$-particle interaction
in the coordinate representation which is defined
through the effective forces $v^{\,(k)}$ entering Eq.~(\ref{dfvk})
and through the generalized antisymmetrized delta functions
by the formulas:
\bea
w^{\,(k)} (x^{\vphantom{'}}_1, \ldots , x^{\vphantom{'}}_k\,;
\, x'_1,\ldots , x'_k) &=& \frac{1}{k\,!\,(k-2)\,!}
\int dx''_1 \cdot \ldots \cdot dx''_k\,
dx'''_1 \cdot \ldots \cdot dx'''_k\,
\nonumber\\
&&\times\;
\delta\! \left( \vphantom{\delta}^{\ds x''_1,\ldots , x''_k}
_{\ds x^{\vphantom{'}}_1, \ldots , x^{\vphantom{'}}_k} \right)\,
\delta\! \left( \vphantom{\delta}^{\ds x'''_1,\ldots , x'''_k}
_{\ds x'_1, \ldots , x'_k} \right)\,
\nonumber\\
&&\times\;
v^{\,(k)} (x''_1, \ldots , x''_k\,;\, x'''_1,\ldots , x'''_k)\,,
\vphantom{\int}
\label{dfwkx}
\eea
where
\be
\delta\! \left( \vphantom{\delta}^{\ds x'_1,\ldots , x'_k}
_{\ds x^{\vphantom{'}}_1, \ldots , x^{\vphantom{'}}_k} \right)
= \det \left(
\begin{array}{ccc}
\delta(x^{\vphantom{'}}_1,\, x'_1) & \ldots &
\delta(x^{\vphantom{'}}_1,\, x'_k) \\
\cdots & \cdots & \cdots \\
\delta(x^{\vphantom{'}}_k,\, x'_1) & \ldots &
\delta(x^{\vphantom{'}}_k,\, x'_k) \\
\end{array}
\right)\,.
\label{gdel}
\ee
Notice that from Eqs. (\ref{dfh0m}), (\ref{prgf1}),
(\ref{eqmgf1}), and (\ref{dfgf0}) it follows:
\be
\Sigma_{\,\xi} (z, z') = -
\Sigma_{\,\xi} (\bar{z}', \bar{z})\,.
\label{prsigm}
\ee
The symmetry property of ${\cal W}^{\,(k)}$ follows from
its definition (\ref{dfwkz}):
\be
{\cal W}^{\,(k)}
(z^{\vphantom{'}}_1, \ldots , z^{\vphantom{'}}_k\,;
\, z'_1,\ldots , z'_k) = (-1)^k\,{\cal W}^{\,(k)}
(\bar{z}'_1,\ldots , \bar{z}'_k\,;
\bar{z}^{\vphantom{'}}_1, \ldots , \bar{z}^{\vphantom{'}}_k)\,.
\label{prwk}
\ee

In order to obtain equations for the other (many-particle)
GFs let us perform a change
of the functional variable $\xi$ to the $G_{\xi}$ and consider
a Legendre transformation of the functional
$W[\,\xi\,]$:
\be
\Gamma [\,G_{\xi}\,] = 2\,W[\,\xi\,]
- \int dz_1\, dz_2\, [\, \xi (z_2, z_1)
- \xi (\bar{z}_1, \bar{z}_2)\,]\,
G_{\xi} (z_1, z_2)\,.
\label{legtr}
\ee
Using Eqs. (\ref{prgf1}) and (\ref{dxi1}) we obtain
\be
\frac{\delta \Gamma}{\delta_{-} G_{\xi} (z_1, z_2)} =
\xi (\bar{z}_1, \bar{z}_2) - \xi (z_2, z_1)\,.
\label{dgamma}
\ee
The notation $\delta_{-}$ means that the variations
of the GF $G_{\xi}$ conserve the property of antisymmetry
(\ref{prgf1}). This condition should be taken into account since
variations of $\xi$, which generate the variations
of $G_{\xi}$, obviously do not lead to the violation of
Eq.~(\ref{prgf1}).
Conservation of the property (\ref{prgf1}) in the variational
procedure can be automatically ensured if the following
substitution is performed in a $G_{\xi}\,$-dependent functional:
\be
G_{\xi}(z, z') = {\txts \frac{1}{2}}
[\,G_{\xi}(z, z') - G_{\xi}(\bar{z}', \bar{z})\,]\,.
\label{gf1as}
\ee
In case of the vanishing source field the Eq.~(\ref{dgamma})
leads to the stationarity condition:
\be
\frac{\delta \Gamma}{\delta_{-} G (z_1, z_2)} = 0\,.
\label{stat}
\ee
Using, further, Eq.~(\ref{eqmgf1}) we obtain from
Eq.~(\ref{dgamma}) the following relation
\bea
\frac{\delta \Sigma_{\,\xi} (z_2, z_1)}
{\delta_{-} G_{\xi} (z_4, z_3)} &=& {\txts \frac{1}{2}}
\left[ \,G^{-1}_{\xi} (z_2, z_4)\, G^{-1}_{\xi} (z_3, z_1)
- G^{-1}_{\xi} (\bar{z}_1, z_4)\, G^{-1}_{\xi} (z_3, \bar{z}_2)
\, \right]
\nonumber\\
&&-\;
\frac{\delta^2 \Gamma}
{\delta_{-} G_{\xi} (z_1, z_2)\,
 \delta_{-} G_{\xi} (z_4, z_3)}\,.
\label{dsigma}
\eea

Let us introduce an amplitude of the effective interaction
${\cal I}_{\xi}$ which includes irreducible amplitudes both
in the particle-hole (ph), and in the particle-particle (pp)
channels:
\be
{\cal I}_{\xi} (z_1, z_2;\,z_3, z_4) =
i\,\frac{\delta \Sigma_{\,\xi} (z_2, z_1)}
{\delta_{-} G_{\xi} (z_4, z_3)}\,.
\label{dfirr}
\ee
From Eqs. (\ref{prgf1}) and (\ref{prsigm}) we obtain:
\be
{\cal I}_{\xi} (z_1, z_2;\,z_3, z_4) = -
{\cal I}_{\xi} (\bar{z}_2, \bar{z}_1;\,z_3, z_4) = -
{\cal I}_{\xi} (z_1, z_2;\,\bar{z}_4, \bar{z}_3)\,.
\label{prpia}
\ee
In addition from Eq.~(\ref{dsigma}) it follows that
\be
{\cal I}_{\xi} (z_1, z_2;\,z_3, z_4) =
{\cal I}_{\xi} (z_4, z_3;\,z_2, z_1)\,.
\label{prpis}
\ee
Notice that the response function defined by Eq.~(\ref{dxi2})
satisfies the analogous equalities
\be
L_{\xi} (z_1, z_2;\,z_3, z_4) = -
L_{\xi} (\bar{z}_2, \bar{z}_1;\,z_3, z_4) = -
L_{\xi} (z_1, z_2;\,\bar{z}_4, \bar{z}_3)\,,
\label{prpla}
\ee
\be
L_{\xi} (z_1, z_2;\,z_3, z_4) =
L_{\xi} (z_4, z_3;\,z_2, z_1)\,.
\label{prpls}
\ee

Differentiating Eq.~(\ref{eqmgf1}) with respect to $\xi$
and then using Eqs. (\ref{dxi2}) and (\ref{dfirr}),
we obtain the Bethe-Salpeter equation (BSE)
for the response function:
\bea
L_{\xi} (z_1, z_2;\,z_3, z_4) &=&
G_{\xi} (\bar{z}_4, z_1)\,G_{\xi} (z_2, \bar{z}_3) -
G_{\xi} (z_3, z_1)\,G_{\xi} (z_2, z_4)
\nonumber\\
&&-\;
i \int dz_5\, dz_6\, dz_7\, dz_8\;
G_{\xi} (z_5, z_1)\,G_{\xi} (z_2, z_6)\,
\nonumber\\
&&\times\;
{\cal I}_{\xi} (z_5, z_6;\,z_7, z_8)\,
L_{\xi} (z_7, z_8;\,z_3, z_4)\,.
\label{bseq}
\eea

The equations for the many-particle GFs $G^{\,(k)}_{\xi}$
with $k>2$ are obtained
by a differentiation of Eq.~(\ref{dfgfkx}) with respect to $\xi$.
Taking into account relation:
\be
\frac{\delta}{\delta \xi (z_1, z_2)} =
\int dz_3\, dz_4\,
L_{\xi}(z_1, z_2;\,z_3, z_4)
\frac{\delta}{\delta G_{\xi} (z_3, z_4)}\,,
\label{dxidg}
\ee
which follows from Eq.~(\ref{dxi2}), we come to the
recurrence formula
\bea
G^{\,(k)}_{\xi}(z^{\vphantom{'}}_1,\ldots ,
z^{\vphantom{'}}_k\,;\, z'_1,\ldots , z'_k) &=&
\biggl[\, G_{\xi} (z^{\vphantom{'}}_1, z'_1)
+ \int dz\, dz'\,
L_{\xi}(z'_1, z^{\vphantom{'}}_1;\,z', z)
\nonumber\\
&&\times\;
\frac{\delta}{\delta G_{\xi} (z', z)}\, \biggr]\;
G^{\,(k-1)}_{\xi}(z^{\vphantom{'}}_2,\ldots ,
z^{\vphantom{'}}_k\,;\, z'_2,\ldots , z'_k)\,.
\label{recgfk}
\eea
Notice that in Eqs.~(\ref{dxidg}) and (\ref{recgfk}) we have:
$\delta_{-} G_{\xi} = \delta G_{\xi}\,$ owing to presence
of the response function satisfying Eqs.~(\ref{prpla}).

The Eqs. (\ref{dfrf}), (\ref{dfsigx}), (\ref{dfirr}),
(\ref{bseq}), and (\ref{recgfk}) form the closed system
of the functional differential equations of the GGFF.
An important feature of these equations is that they do not
change their form when the many-particle forces are added to
the two-particle interaction in the total Hamiltonian.
The exception is Eq.~(\ref{dfsigx}) for the mass operator
in which the many-particle forces enter in the explicit form.
This result of the GGFF could be expected, but it needed to be
proved. It allows to extend the standard GF methods developed
for the Fermi systems with two-particle interaction to the
systems interacting through the many-particle effective forces.

In the final equations we can set $\xi=0$.
So, in what follows we shall omit the
$\xi$-indices of the functions implying the limit at $\xi=0$
and coming back to the GFs without source field
defined by Eq.~(\ref{dfgfk}). The independent functional variable
is the single-particle GF $G$ satisfying the Dyson equation
which follows from Eq.~(\ref{eqmgf1}):
\be
G (z_1, z_2) = G^0 (z_1, z_2) + \int dz_3\, dz_4\,
G^0 (z_1, z_3)\,\Sigma (z_3, z_4)\,G (z_4, z_2)\,.
\label{dyson}
\ee
In case of the Fermi systems without pairing correlations,
when the external anomalous pair potentials
vanish in Eq.~(\ref{dfh0}), the above equations can be reduced
to ones for the normal components of GFs which satisfy the
condition:
\be
\sum_{n=1}^k (\chi^{\vphantom{'}}_n - \chi'_n)\,
G^{\,(k)} (z^{\vphantom{'}}_1,\ldots , z^{\vphantom{'}}_k\,;\,
z'_1,\ldots , z'_k) = 0\,.
\label{gfknor}
\ee
These reduced equations will
coincide with ones obtained in Ref.~\cite{T93}.
On the other hand, if we restrict consideration of the Fermi
systems with pairing to the case of
two-particle interaction in the Hamiltonian
(\ref{dfhtot}), i.e. if we assume:
\be
H=H^0 + V^{\,(2)}\,,
\label{dfhtot2}
\ee
the Eqs. (\ref{dfrf}), (\ref{dfsigx}), (\ref{dfirr}), and
(\ref{bseq}) will coincide up to the permutation of the
arguments with the corresponding equations in Ref.~\cite{IS77}.

\subsection{Transformation of the basic equations \label{sec23}}

It is easy to see that any amplitude $\cal I$
possessing the properties (\ref{prpia}) and (\ref{prpis})
can be represented in the form
\be
{\cal I} (z_1, z_2;\,z_3, z_4) = {\txts \frac{1}{2}}\,
[\,
{\cal U} (z_1, z_2;\,z_3, z_4) -
{\cal U} (\bar{z}_2, \bar{z}_1;\,z_3, z_4)\,]\,,
\label{dfu1}
\ee
where the amplitude $\cal U$ satisfies the equalities
\be
{\cal U} (z_1, z_2;\,z_3, z_4) =
{\cal U} (z_4, z_3;\,z_2, z_1) =
{\cal U} (\bar{z}_2, \bar{z}_1;\,\bar{z}_4, \bar{z}_3)\,,
\label{prpus}
\ee
but generally $\cal U$ does not possess the property of antisymmetry
(\ref{prpia}). So in the following we shall refer to the $\cal U$
as the non-antisymmetric amplitude of the effective interaction.
Obviously, the Eq.~(\ref{dfu1}) does not define $\cal U$ uniquely.
To specify the definition let us note that Eqs.~(\ref{prpia}) are
fulfilled if the functional variable of $\Sigma_{\,\xi}$ in
Eq.~(\ref{dfirr}) is taken in the form (\ref{gf1as}). However,
one can formally define a functional derivative in which the
substitution (\ref{gf1as}) is not supposed. Following this definition
we set:
\be
{\cal U} (z_1, z_2;\,z_3, z_4) =
i\,\frac{\delta \Sigma (z_2, z_1)}{\delta G (z_4, z_3)}\,,
\label{dfu2}
\ee
where the condition (\ref{prgf1}) is not supposed to be
fulfilled under variations of $G$.

As an example, let us consider the first order in the
two-particle interaction. From Eqs. (\ref{dfsigx}),
(\ref{dfrf}), and (\ref{bseq}) we have:
\be
\Sigma^{\,[1]} (z_1, z_2) = i \int dz_3\, dz_4\,[\,
{\cal W}^{\,(2)} (z_1, z_4;\,z_3, z_2)
- {\txts \frac{1}{2}}\,
{\cal W}^{\,(2)} (z_1, \bar{z}_2;\,z_3, \bar{z}_4)\,]
\,G (z_3, z_4)\,.
\label{sig1}
\ee
$\Sigma^{\,[1]}$ is the generalized Hartree-Fock-Bogoliubov (HFB)
contribution into the mass operator
[the term {\sl generalized} is used since
the exact single-particle GF enters Eq.~(\ref{sig1})].
Using, further, the definitions (\ref{dfu2}) and (\ref{dfirr})
we obtain from Eq.~(\ref{sig1}):
\bea
{\cal U}^{\,[1]} (z_1, z_2;\,z_3, z_4) &=&
{\cal W}^{\,(2)} (z_2, z_3;\,z_1, z_4)
+ {\txts \frac{1}{2}}\,
{\cal W}^{\,(2)} (z_2, \bar{z}_1;\,z_4, \bar{z}_3)\,,
\label{u1}\\
{\cal I}^{\,[1]} (z_1, z_2;\,z_3, z_4) &=&
{\txts \frac{1}{2}}\,[\,
{\cal U}^{\,[1]} (z_1, z_2;\,z_3, z_4) -
{\cal U}^{\,[1]} (\bar{z}_2, \bar{z}_1;\,z_3, z_4)\,]\,.
\label{irr1}
\eea
It can be seen that Eqs.~(\ref{prpus}) for the ${\cal U}^{\,[1]}$
follow from the antisymmetry of the ${\cal W}^{\,(k)}$
and from the property (\ref{prwk}), while Eqs.~(\ref{prpia})
are not fulfilled.

For the further applications it is convenient to
introduce a non-antisymmetric response function $R$
satisfying the following BSE:
\bea
R (z_1, z_2;\,z_3, z_4) &=&
R^{\,0} (z_1, z_2;\,z_3, z_4)
+ i \int dz_5\, dz_6\, dz_7\, dz_8\;
\nonumber\\
&&\times\;
R^{\,0} (z_1, z_2;\,z_5, z_6)\,
{\cal U} (z_5, z_6;\,z_7, z_8)\,
R (z_7, z_8;\,z_3, z_4)\,,
\label{bseqr}
\eea
where
\be
R^{\,0} (z_1, z_2;\,z_3, z_4)=
-G (z_3, z_1)\,G (z_2, z_4)\,.
\label{dfr0}
\ee
It is easy to show, first that the following equalities hold:
\be
R (z_1, z_2;\,z_3, z_4) =
R (z_4, z_3;\,z_2, z_1) =
R (\bar{z}_2, \bar{z}_1;\,\bar{z}_4, \bar{z}_3)\,,
\label{prprs}
\ee
while the Eqs.~(\ref{prpla}) do not hold for $R$.
Second, the response function $L$, which
satisfies Eq.~(\ref{bseq}), is expressed in terms of $R$ as:
\be
L (z_1, z_2;\,z_3, z_4)=
R (z_1, z_2;\,z_3, z_4)-
R (\bar{z}_2, \bar{z}_1;\,z_3, z_4)\,.
\label{lrmr}
\ee
Thus, for the determination of the response function $L$
it is sufficient to solve Eq.~(\ref{bseqr}) for the
non-antisymmetric function $R$.

Now, following the method described in Ref.~\cite{KTT},
we represent the total mass operator $\Sigma$ and the total
non-antisymmetric amplitude of the effective interaction $\cal U$
as a sum of two terms:
\be
\Sigma = \tilde{\Sigma} + \Sigma^{\,e}\,,\qquad
{\cal U} = \tilde{\cal U} + {\cal U}^{\,e}\,,
\label{div}
\ee
where
\bea
\tilde{\Sigma} (z_1, z_2) &=&
\tilde{\Sigma} (y_1, y_2)\;\delta (t_1 - t_2)\,,
\label{tsgt}\\
\tilde{\cal U} (z_1, z_2;\,z_3, z_4) &=&
\tilde{\cal U} (y_1, y_2;\,y_3, y_4)\;
\delta (t_1 - t_2)\;\delta (t_3 - t_4)\;\delta (t_1 - t_3)\,.
\label{tut}
\eea
After transformation to the energy representation
(see Sec.~\ref{sec24})
the first terms in Eqs.~(\ref{div}), $\tilde{\Sigma}$
and $\tilde{\cal U}$, are found to be energy-independent.
The term $\tilde{\Sigma}$ corresponds to the mean-field
contribution into the mass operator including the pair
potentials. The term $\tilde{\cal U}$ corresponds to the
residual energy-independent interaction both in the ph,
and in the pp channels. The second terms in
Eqs.~(\ref{div}), $\Sigma^{\,e}$ and ${\cal U}^{\,e}$,
have a strong energy dependence and represent
dynamic contributions of complex configurations.

We stress that the quantities $\tilde{\Sigma}$
and $\tilde{\cal U}$ (and, consequently,
$\Sigma^{\,e}$ and ${\cal U}^{\,e}$) are not defined
rigorously by Eqs.~(\ref{div}). They will be specified
in the following within the framework of the model
to be considered. At the moment only the general properties,
which are expressed by Eqs. (\ref{tsgt}) and (\ref{tut}),
are important. Notice, however, that in particular case
of the self-consistent HFB approximation restricted by
the two-particle interaction we have:
$\tilde{\Sigma}=\Sigma^{\,[1]}$,
$\;\tilde{\cal U}={\cal U}^{\,[1]}$, where the
right-hand sides are defined by Eqs. (\ref{sig1}) and
(\ref{u1}) with the exact GF $G$ being replaced by the
HFB GF $\tilde{G}$ which is the solution of Eq.~(\ref{dyson2})
(see below).

Using decompositions (\ref{div}) one can transform both of
Eqs. (\ref{dyson}) and (\ref{bseqr}) to the system of two
equations. In the symbolic notations we have:
\bea
G &=& \tilde{G} + \tilde{G}\,\Sigma^{\,e}\,G\,,
\label{dyson1}\\
\tilde{G} &=& G^0 + G^0\,\tilde{\Sigma}\,\tilde{G}\,,
\label{dyson2}\\
R &=& R^{\,e} + i\,R^{\,e}\,\tilde{\cal U}\,R\,,
\label{bseqr1}\\
R^{\,e} &=& R^{\,0} + i\,R^{\,0}\,{\cal U}^{\,e}\,R^{\,e}\,.
\label{bseqr2}
\eea
Proceeding by the same method as in Refs.~\cite{KTT,T89}
the last equation can be brought to the following form:
\bea
R^{\,e} (z_1, z_2;\,z_3, z_4)\!\!&=&\!\!
\tilde{R}^{\,0} (z_1, z_2;\,z_3, z_4)
+ i \int dz_5\, dz_6\, dz_7\, dz_8
\nonumber\\
&&\!\!\times\;
\tilde{R}^{\,0} (z_1, z_2;\,z_5, z_6)\,
{\cal W}^{\,e} (z_5, z_6;\,z_7, z_8)\,
R^{\,e} (z_7, z_8;\,z_3, z_4)\,,
\label{bseqr3}
\eea
where
\bea
\tilde{R}^{\,0} (z_1, z_2;\,z_3, z_4) &=&
-\tilde{G} (z_3, z_1)\,\tilde{G} (z_2, z_4)\,,
\label{dfr0t}\\
{\cal W}^{\,e} (z_1, z_2;\,z_3, z_4) &=&
{\cal V}^{\,e} (z_1, z_2;\,z_3, z_4) -
i\,\Sigma^{\,e} (z_3, z_1)\,\Sigma^{\,e} (z_2, z_4)\,,
\label{dfwe}\\
{\cal V}^{\,e} (z_1, z_2;\,z_3, z_4) &=&
{\cal U}^{\,e} (z_1, z_2;\,z_3, z_4)
\nonumber\\
&&+\;
i\,\Sigma^{\,e} (z_3, z_1)\,\tilde{G}^{-1} (z_2, z_4) +
i\,\tilde{G}^{-1} (z_3, z_1)\,\Sigma^{\,e} (z_2, z_4)\,.
\label{dfwet}
\eea
The Eq.~(\ref{bseqr3}) is the basic one for the building of
model which will be considered in the second part of the paper.

\subsection{Single-quasiparticle basis functions
            and the energy representation \label{sec24}}

For the following analysis it is required to introduce
a set of basis functions
$\{\psi^{\vphantom{*}}_{\vphantom{1'}1}(y)\}$
in the extended space defined previously in Sec.~\ref{sec21}.
The usual conditions of orthonormality and completeness are
supposed to be fulfilled:
\be
\int dy\,
\psi^*_{\vphantom{1'}1}(y)\,
\psi^{\vphantom{*}}_{1'}(y) =
\delta_{1,1'}\,,
\qquad
\sum_1\,
\psi^*_{\vphantom{1'}1}(y)\,
\psi^{\vphantom{*}}_{\vphantom{1'}1}(y') =
\delta (y,y')\,.
\label{ortcom}
\ee
It is convenient to consider
$\psi^{\vphantom{*}}_{\vphantom{1'}1}(y)$
to be the eigenfunctions of the operator:
\be
{\cal H} (y, y') = {\cal H}^0 (y, y') + \tilde{\Sigma} (y, y')\,,
\label{dfch}
\ee
where ${\cal H}^0$ defines the single-particle term of the
total Hamiltonian according to
Eqs. (\ref{dfh0b}) and  (\ref{dfh0m}),
$\tilde{\Sigma}$ is the mean-field contribution into the
total mass operator in Eqs.~(\ref{div}).
Thus, we assume the following equation to be fulfilled:
\be
\int dy'\,{\cal H} (y, y')\,
\psi^{\vphantom{*}}_{\vphantom{1'}1}(y') =
E^{\vphantom{*}}_{\vphantom{1'}\,1}
\psi^{\vphantom{*}}_{\vphantom{1'}1}(y)\,.
\label{psieq}
\ee
Since $\cal H$ possesses the same symmetry properties as the
operators ${\cal H}^0$ and $\tilde{\Sigma}$, i.e.:
\be
{\cal H} (y, y') = {\cal H}^* (y', y) = -
{\cal H} (\bar{y}', \bar{y})\,,
\label{prpch}
\ee
it is not difficult to see that
the complete set of the eigenfunctions of $\cal H$
is divided into two equal parts
which are related by the operation of conjugation:
\be
\psi^{\vphantom{*}}_{\bar{1}}(y) =
\psi^*_{\vphantom{\bar{1}}1}(\bar{y})\,.
\label{conj}
\ee
For the corresponding eigenvalues we have:
$E_{\,\bar{1}} = - E_{\vphantom{\bar{1}}\,1}$.
So one can denote:
$1=\{\lambda_1,\eta_1\}$, $\bar{1}=\{\lambda_1,-\eta_1\}$,
where $\lambda_1$ is the index of the usual single-particle
configuration space (e.g.,
$\lambda = \{\tau_{\lambda}, n, l, j, m \}$
for the spherically symmetric system), $\eta_1 = \pm 1$
is the sign of the eigenvalue $E_{\vphantom{\bar{1}}\,1}$:
\be
E_{\vphantom{\bar{1}}\,1} = \eta_1 E_{\lambda_1}\,,\qquad
E_{\lambda_1} = |E_{\vphantom{\bar{1}}\,1}|\,.
\label{eigv}
\ee

In the representation of $\psi$-functions, the operators
$b$ defined by Eq.~(\ref{dfb}) have the form
\be
b^{\vphantom{\dag}}_1=\int dy\,\psi^*_1(y)\,b(y)\,.
\label{dfb1}
\ee
It is worth noting that the $b$-operators in this
representation are simply related to the
creation and annihilation operators of the
quasiparticles $\alpha^{\dag}_{\lambda}$ and
$\alpha^{\vphantom{\dag}}_{\lambda}$ which are usually introduced
in the HFB theory. Namely, we have (see Ref.~\cite{KSTV}):
$b^{\vphantom{\dag}}_{\lambda,+}=
\alpha^{\vphantom{\dag}}_{\lambda}$,
$b^{\vphantom{\dag}}_{\lambda,-}=\alpha^{\dag}_{\lambda}$.
So, in what follows we shall refer to
the functions $\psi^{\vphantom{*}}_{\vphantom{1'}1}(y)$
as the single-quasiparticle functions.

A more detailed form of the functions
$\psi^{\vphantom{*}}_{\vphantom{1'}1}(y)$ can be obtained
making use of the Bloch-Messiah theorem, see Refs.~\cite{RS,BM}.
To formulate the result let us note that for
time-reversal invariant fermion system,
the set of the single-particle indices
$\lambda$ can be divided into two subsets of conjugate indices
$p$ and $\bar{p}$ which represent ``paired'' states, i.e.:
$\{ \lambda \} = \{ p \} \cup \{ \bar{p} \}$.
In particular, for spherically symmetric system
the conjugate indices are $\lambda = \{(\lambda), m \}$
and $\bar{\lambda} = \{(\lambda), -m \}$ where
$(\lambda) = \{\tau_{\lambda}, n, l, j \}$.
According to the Bloch-Messiah theorem the single-quasiparticle
functions can be represented in the following form
(see Ref.~\cite{KSTV}):
\be
\psi^{\vphantom{*}}_{\lambda , +}(y) = \sum_{\lambda'}
C^{\vphantom{*}}_{\lambda'\lambda}\,
\check{\psi}^{\vphantom{*}}_{\lambda' , +}(y)\,, \qquad
\psi^{\vphantom{*}}_{\lambda , -}(y) = \sum_{\lambda'}
C^*_{\lambda'\lambda}\,
\check{\psi}^{\vphantom{*}}_{\lambda' , -}(y)\,,
\label{unitr}
\ee
where $C^{\vphantom{*}}_{\lambda'\lambda}$ is a unitary matrix and
the functions $\check{\psi}_{\lambda , \eta}(y)$ have the form:
\be
\left.
\begin{array}{rclrcl}
\check{\psi}_{p\,,+}(x,+) &=& \hphantom{-}
u^{\vphantom{*}}_{p}\,
\vphi^{\vphantom{*}}_{p}(x)\,,\qquad&
\check{\psi}_{\bar{p}\,,+}(x,+) &=&
u^{\vphantom{*}}_{p}\,
\vphi^{\vphantom{*}}_{\bar{p}}(x)\,,\\
\check{\psi}_{p\,,+}(x,-) &=&
-v^{\vphantom{*}}_{p}\,
\vphi^*_{\bar{p}}(x)\,,\qquad&
\check{\psi}_{\bar{p}\,,+}(x,-) &=&
v^{\vphantom{*}}_{p}\,
\vphi^*_{p}(x)\,,\\
\check{\psi}_{p\,,-}(x,+) &=&
-v^{\vphantom{*}}_{p}\,
\vphi^{\vphantom{*}}_{\bar{p}}(x)\,,\qquad&
\check{\psi}_{\bar{p}\,,-}(x,+) &=&
v^{\vphantom{*}}_{p}\,
\vphi^{\vphantom{*}}_{p}(x)\,,\\
\check{\psi}_{p\,,-}(x,-) &=& \hphantom{-}
u^{\vphantom{*}}_{p}\,
\vphi^*_{p}(x)\,,\qquad&
\check{\psi}_{\bar{p}\,,-}(x,-) &=&
u^{\vphantom{*}}_{p}\,
\vphi^*_{\bar{p}}(x)\,.\\
\end{array}
\right\}
\label{uvpsi}
\ee
Here $\{ \vphi_{\lambda}(x) \}$ is a complete set of
orthonormal functions in the usual single-particle space,
$v_{\lambda}$ and $u_{\lambda}$ are real non-negative
numbers which satisfy the following conditions:
$\,u^{\vphantom{2}}_{\lambda} =
\sqrt{1 - v^2_{\lambda} \vphantom{V^A}}\,$,
$\,v_{p} = v_{\bar{p}} \leqslant 1$.
For the spherically symmetric system we have:
\be
\left.
\begin{array}{rcl}
\check{\psi}_{\lambda,+}(x,+) = u^{\vphantom{*}}_{\lambda}\,
\vphi^{\vphantom{*}}_{\lambda}(x)\,,\quad &
\check{\psi}_{\lambda,+}(x,-) = (-1)^{l+j+m}\,
v^{\vphantom{*}}_{\lambda}\, \vphi^*_{\bar{\lambda}}(x)\,,\\
\check{\psi}_{\lambda,-}(x,-) = u^{\vphantom{*}}_{\lambda}\,
\vphi^*_{\lambda}(x)\,,\quad &
\check{\psi}_{\lambda,-}(x,+) = (-1)^{l+j+m}\,
v^{\vphantom{*}}_{\lambda}\,
\vphi^{\vphantom{*}}_{\bar{\lambda}}(x)\,.\\
\end{array}
\right\}
\label{uvpsis}
\ee

Up to now we did not restrict our analysis to the systems
where the number of particles is conserved exactly.
The reason is that our aim was modification
of the existing general GF formalism for the arbitrary
Fermi systems with pairing.
However, application of the formalism to the atomic nuclei,
we are interested in, implies that
the particle-number conservation law is fulfilled.
Thus, in what follows we assume that the total Hamiltonian $H$
defined by Eqs. (\ref{dfhtot})--(\ref{dfvk}) does not contain
the external anomalous pair potentials $\Delta^0$
and that Eqs.~(\ref{dfhd0}) hold.
In that case the ground-state wave function
$|\, 0 \,\rangle$ which enters the definition of the GFs
(\ref{dfgfk}) is an eigenfunction of the particle-number
operator. This means that the exact GFs do not contain
anomalous components. In particular, the exact single-particle
GF satisfies condition [cf. Eq.~(\ref{gfknor})]:
\be
(\chi - \chi')\,G(z,z')=0\,.
\label{anzer}
\ee
However, this condition is not
fulfilled for the GF $\tilde{G}$ which is the solution of
Eq.~(\ref{dyson2}) with the operator $\tilde{\Sigma}$
including the pair potentials independently of the Hamiltonian
$H$ (e.g., within the HFB approximation). Justification of using
such GF $\tilde{G}$ is as follows. It enables one to take into
account pairing correlations effectively and should be
considered only as an approximation to the exact GF.
The latter is found from Eq.~(\ref{dyson1}) in which the
mass operator $\Sigma^{\,e}$ has to contain all necessary
corrections to $\tilde{\Sigma}$, such that the solution of
Eq.~(\ref{dyson1}) satisfies Eq.~(\ref{anzer}).
Of course this scheme should be considered only as
a philosophy of the approach, i.e. as an ideal program
which is difficult to implement completely in practice.

Let us now define the energy representation of the
Green functions and of the related quantities entering
above equations. Making use of the basis
$\{\psi^{\vphantom{*}}_{\vphantom{1'}1}(y)\}$,
let us introduce the following Fourier transformations:
\bea
G_{12} (\ve) &=&
\int dz_1\,dz_2\;
\psi^*_1(y_1)\,\psi^{\vphantom{*}}_2(y_2)\,
\delta(t_2)\,
\exp\bigl(\,i\ve\,(t_1-t_2)\,\bigr)\,
G (z_1, z_2)\,,
\label{gf1e}\\
R_{12,34} (\omega) &=& - i
\int dz_1\,dz_2\,dz_3\,dz_4\;
\psi^{\vphantom{*}}_1(y_1)\,\psi^*_2(y_2)\,\psi^*_3(y_3)\,
\psi^{\vphantom{*}}_4(y_4)
\nonumber\\
&&\times\;
\delta(t_1-t_2-0)\,\delta(t_4-t_3-0)\,\delta(t_4)\,
\exp\bigl(\,i\omega\,(t_3-t_1)\,\bigr)
\nonumber\\
&&\times\;
R (z_1, z_2;\,z_3, z_4)\,,\vphantom{\int}
\label{rfe}\\
{\cal U}_{\,12,34} (\omega,\,\ve,\,\ve') &=&
\int dz_1\,dz_2\,dz_3\,dz_4\;
\psi^{\vphantom{*}}_1(y_1)\,\psi^*_2(y_2)\,\psi^*_3(y_3)\,
\psi^{\vphantom{*}}_4(y_4)
\nonumber\\
&&\times\;
\delta(t_4)\,\exp\bigl(\,i\omega\,(t_3-t_1) +
i\ve\,(t_2-t_1) + i\ve'\,(t_3-t_4)\,\bigr)
\nonumber\\
&&\times\;
{\cal U} (z_1, z_2;\,z_3, z_4)\,.\vphantom{\int}
\label{utote}
\eea
The quantities $\tilde{G}_{12}(\ve)$,
$\Sigma^{\vphantom{e}}_{\,12}(\ve)$, $\Sigma^{\,e}_{\,12}(\ve)$,
$L^{\vphantom{e}}_{12,34} (\omega)$,
${\cal U}^{\,e}_{\,12,34} (\omega,\,\ve,\,\ve')$,
and others are defined in analogy to these formulas.
In accordance with Eqs. (\ref{lrmr}), (\ref{conj}), and
(\ref{rfe}) we have:
\be
L_{12,34} (\omega) = R_{12,34} (\omega) -
R_{\bar{2}\bar{1},34} (\omega)\,.
\label{lrmre}
\ee
Notice that spectral expansion for the response function
$L(\omega)$ has the form:
\be
L_{12,34} (\omega) = - \sum_{\eta = \pm 1} \sum_{n \ne 0}
\frac{\eta\,\rho^{n(\eta)}_{12}\,\rho^{n(\eta)^{\ds *}}_{34}}
{\,\omega - \eta\,(\,\omega_n - i\!\cdot\!0)}\,,
\label{spectl}
\ee
which is similar to the analogous formula for the Fermi systems
without pairing correlations (see, e.g., Ref~\cite{IS74}).
The difference consists in the definition of the transition
amplitudes. In Eq.~(\ref{spectl}) we have:
\be
\rho^{n(\eta)}_{12} =
\delta^{\vphantom{(+)}}_{\eta,+1}\,
\langle\, n\, |\,b^{\dag}_1\,b^{\vphantom{\dag}}_2\,
|\, 0 \,\rangle +
\delta^{\vphantom{(+)}}_{\eta,-1}\,
\langle\, 0\, |\,b^{\dag}_1\,b^{\vphantom{\dag}}_2\,
|\, n \,\rangle \,,\qquad
\omega_n = E_n - E_0\,,
\label{tran}
\ee
where $|\, n \,\rangle$, $|\, 0 \,\rangle$, $E_n$, and $E_0$
are the eigenfunctions and the eigenvalues of the
Hamiltonian $H$. Taking into account the definitions
(\ref{dfb}) and (\ref{dfb1}), one can see
that even if the ground-state wave function
$|\, 0 \,\rangle$ is an eigenfunction of the particle-number
operator, the amplitudes $\rho^{n(\eta)}_{12}$ take nonzero
values not only for the transitions between the states
with the same number of particles,
but also for the transitions between the
ground state of the $N$-particle system $|\, 0 \,\rangle$ and
the states $|\, n \,\rangle$ of the systems consisting of
$N \pm 2$ particles. Thus, the spectral expansion (\ref{spectl})
contains information about excitations of the $N$-particle
system both in the ph, and in the pp channels.

\section{QUASIPARTICLE TIME BLOCKING APPROXIMATION
         \label{sec3}}

\subsection{General framework \label{sec31}}

Let us now turn to the question of determining the physical
observables and related quantities in this approach, namely,
excitation energies $\omega_n$ and transition amplitudes
$\rho^{n(\eta)}_{12}$. It follows from the Eqs. (\ref{lrmre})
and (\ref{spectl}) that in order to find these characteristics
we need to know the response function $R(\omega)$, i.e. to
solve the system of equations (\ref{bseqr1}) and (\ref{bseqr3}).
The basic difficulty in solving this task is that the
Eq.~(\ref{bseqr3}) contains energy-dependent interaction
${\cal U}^{\,e}$ and mass operator $\Sigma^{\,e}$.
Notice, however, that these energy-dependent quantities
arise only in the case when the dynamic contributions
of complex configurations are taken into account explicitly.
In the energy representation Eq.~(\ref{bseqr3}) is an integral
equation for the function $R^{\,e}(\omega,\,\ve,\,\ve')$
[defined in analogy to Eq.~(\ref{utote})] over the energy
variable $\ve$, which cannot be, strictly speaking, reduced to
the closed equation for $R^{\,e}(\omega)$ due to the energy
dependence of ${\cal U}^{\,e}$ and $\Sigma^{\,e}$.
Fortunately, there are methods that allow us to avoid
complicated problem of exact solution of this equation,
making use of certain approximations.
One of such methods will be considered here.

We begin by noting that if we use the eigenfunctions of
the operator ${\cal H}$, i.e. the set
$\{\psi^{\vphantom{*}}_{\vphantom{1'}1}(y)\}$,
as the basis functions, the single-particle GF
$\tilde{G}$ is diagonal:
\be
\tilde{G}_{12}(\ve) = \frac{\delta_{12}}
{\ve - E_1 + i\,\eta_1\!\cdot\!0}\;,
\label{tgf1e}
\ee
as follows from Eqs. (\ref{dfgf0}), (\ref{dyson2}),
(\ref{dfch}), and (\ref{psieq}).
In the time representation we have:
\be
\tilde{G}_{12}(t_1,t_2) = -i\,\eta_1 \delta_{12}\,
\theta (\eta_1 t_{12})\,\exp(-i E_1 t_{12})\,,
\label{tgf1t}
\ee
where $t_{12}=t_1-t_2$, $\,\theta (\tau)$ is the step function.
These expressions are formally identical with analogous
formulas for the normal GFs, except that they are written
in the extended basis representation.
It enables one to apply, practically without changes,
the method of chronological decoupling of diagrams
(MCDD) to the solution of Eqs. (\ref{bseqr1}) and
(\ref{bseqr3}) which contain the GFs with pairing.
The MCDD was developed in Ref.~\cite{T89} for the solution
of the ph-channel BSE in the normal Fermi system
including dynamic effects both in the interaction
and in the mass operator.
The idea of the method is similar to that used
in the other methods developed earlier for the solution
of the analogous problems in Refs. \cite{Wu} (ph-channel BSE)
and \cite{HH75,HEH77,HK89} (pp-channel BSE).
However, the MCDD differs from the aforementioned methods
in some details, in particular concerning the treatment
of the GSC2.
Almost all the resulting equations obtained by means of
this straightforward extension of the MCDD are found to be
formally identical with equations for the normal Fermi system
in the same sense as Eqs. (\ref{tgf1e}) and (\ref{tgf1t}).
Because derivation of these equations in the latter case
was described in detail in Refs.~\cite{KTT,T89},
we will draw only the main formulas and the final results.

First of all, the function $\tilde{R}^{\,0}=-\tilde{G}\tilde{G}$
entering Eq.~(\ref{bseqr3}) is divided into two parts:
$\tilde{R}^{\,0}=\tilde{R}^{\,0(a)}+\tilde{R}^{\,0(b)}$
where
\be
\tilde{R}^{\,0(a)}_{12,34}(t_1,t_2\,;\;t_3,t_4) =
-\delta^{\vphantom{(+)}}_{\eta_{\mbts{1}},-\eta_{\mbts{2}}}\,
\theta (\eta_1 t_{41})\,\theta (\eta_1 t_{32})\,
\tilde{G}_{31}(t_3,t_1)\,\tilde{G}_{24}(t_2,t_4)\,,
\label{trza}
\ee
$\tilde{R}^{\,0(b)}$ is the remainder term,
which is absorbed in part in the renormalization procedure.
As compared with the initial function $\tilde{R}^{\,0}$
defined by Eq.~(\ref{dfr0t}),
the term $\tilde{R}^{\,0(a)}$ contains two additional
time-dependent step functions and the factor
$\delta^{\vphantom{(+)}}_{\eta_{\mbts{1}},-\eta_{\mbts{2}}}$
which play twofold role.
On the one hand, they allow to obtain closed set of the
algebraic equations in the energy representation
for the main component of the function $R^{\,e}(\omega)$
[see Eqs. (\ref{aplusb}), (\ref{dfprop}), (\ref{ammeq}) below]
which is much more simple for the solution as compared
with the initial Eq.~(\ref{bseqr3}).
On the other hand, owing to these additional
$\theta$ and $\delta$ functions, an approximate
solution of the Eq.~(\ref{bseqr3}) obtained
by this way contains the contributions of the 2q
and 2q$\otimes$phonon configurations, while
more complicated intermediate states
(e.g., 2q$\otimes$2phonon, 2q$\otimes$3phonon, and so on)
are blocked, in part, in the time representation.
So in the following this scheme will be referred to as
the quasiparticle time blocking approximation (QTBA).

Further, renormalization procedure is applied to
Eq.~(\ref{bseqr1}) for the response function $R$, which leads
to the following equation for the effective response function
$R^{\,\mbsu{eff}}$ in the energy representation:
\be
R^{\,\mbsu{eff}}_{12,34} (\omega) =
A^{\vphantom{\mbsu{eff}}}_{12,34} (\omega) -
\sum_{5678} A^{\vphantom{\mbsu{eff}}}_{12,56} (\omega)\,
{\cal F}^{\vphantom{\mbsu{eff}}}_{56,78}\,
R^{\,\mbsu{eff}}_{78,34} (\omega)\,,
\label{bseqren}
\ee
where $A(\omega)$ is a joint (ph and pp) correlated
propagator, ${\cal F}$ is an amplitude of the
effective interaction. The propagator $A(\omega)$ is the main
term of the formal decomposition
\be
R^{\,e}_{12,34} (\omega) = A_{12,34} (\omega) + B_{12,34}\,.
\label{aplusb}
\ee
It contains: (i) the sum of an infinite number of terms
to all orders in $\tilde{R}^{\,0(a)}$, and
(ii) the terms linear and quadratic in $\tilde{R}^{\,0(b)}$
which are related to the GSC2.
The term $B$ in Eq.~(\ref{aplusb})
is an auxiliary quantity which is supposed to be
energy-independent. In addition, it is supposed that $B$
contains all the contributions which are not included
explicitly in the propagator $A(\omega)$. The effective
interaction ${\cal F}$ and the effective charge operator $e$
are defined by the formulas:
\be
{\cal F}^{\vphantom{\dag}}_{12,34} = \sum_{56}
e^{\vphantom{\dag}}_{12,56}\,
\tilde{{\cal U}}^{\vphantom{\dag}}_{\,56,34}\,,
\qquad
e^{\vphantom{\dag}}_{12,34} =
\delta^{\vphantom{\dag}}_{13}\,\delta^{\vphantom{\dag}}_{24}
- \sum_{56}
{\cal F}^{\vphantom{\dag}}_{12,56}\,
B^{\vphantom{\dag}}_{56,34} =
(e^{\,\dag}_{34,12})^{\ds *}\,.
\label{dfefch}
\ee
In terms of these quantities the exact response function $R$
is related to the effective response function by the ansatz:
\be
R^{\vphantom{*}}_{12,34} (\omega) = \sum_{5678}
e^{\,\dag}_{12,56}
\,R^{\,\mbsu{eff}}_{56,78} (\omega)\,
e^{\vphantom{\dag}}_{78,34} +
\sum_{56} B^{\vphantom{*}}_{12,56}\,
e^{\vphantom{\dag}}_{56,34}\,.
\label{rrren}
\ee

One of the basic quantities, which determines the physical
observables in this approach, is the nuclear polarizability
$\Pi (\omega)$. More precisely, it determines distribution of
the transition strength caused by an external field
$V^{\,0}(x,x')$. The function $\Pi (\omega)$ is defined as
\be
\Pi (\omega) = - \frac{1}{2}\,\sum_{1234}\,
(e\,V^{\,0})^{\ds *}_{21}
\,R^{\,\mbsu{eff}}_{12,34} (\omega)\,
(e\,V^{\,0})^{\vphantom{\ds *}}_{43}\,,
\qquad
(e\,V^{\,0})^{\vphantom{*}}_{12}
= \sum_{34}
e^{\vphantom{\dag}^{\vphantom{\ds *}}}_{21,43}\,
V^{\,0}_{\,34}\,,
\label{dfpol1}
\ee
where
\be
V^{\,0}_{\,\vphantom{\bar{1}}12} = \int dy\,dy'\,
\psi^*_1(y)\,\psi^{\vphantom{*}}_2(y')\,
\delta_{\chi,\chi'}\,
\bigl[\,\delta_{\chi,+1}\,V^{\,0}(x,x')
- \delta_{\chi,-1}\,V^{\,0}(x',x)\,\bigr] =
- V^{\,0}_{\,\bar{2}\bar{1}}\,.
\label{dfextv}
\ee
In particular, the strength function $S(E)$ which is frequently
used for the description of nuclear excitations
is expressed in terms of the polarizability as
\be
S(E) = \frac{1}{2\pi}\,\mbox{Im}\,\sum_{1234}\,
V^{\,0^{\ds *}}_{\,21}
\,R^{\vphantom{0}}_{12,34} (E + i\,\Delta)\,
V^{\,0}_{\,43} =
- \frac{1}{\pi}\,\mbox{Im}\,\Pi(E + i\,\Delta)\,,
\label{dfstrf}
\ee
where $\Delta$ is a smearing parameter.
The formulas (\ref{bseqren})--(\ref{dfpol1}), (\ref{dfstrf})
are completely analogous to the ones for
the normal Fermi system (see Ref.~\cite{KTT}), except for
the factors $\frac{1}{2}$ in Eqs. (\ref{dfpol1}) and
(\ref{dfstrf}) which arise
due to definition (\ref{dfextv}) of the operator
$V^{\,0}$ in the extended space taken in the
antisymmetric form.

\subsection{Correlated propagator within the QTBA
            \label{sec32}}

Eq.~(\ref{bseqren}) for the response function is still quite
general. To formulate a model we have to define the correlated
propagator $A(\omega)$. In particular, if we neglect the dynamic
contributions of complex configurations, i.e. if we put
${\cal U}^{\,e}=0$ and $\Sigma^{\,e}=0$, we come to the QRPA.
In this case we have: $A(\omega)=\tilde{A}(\omega)$ where
$\tilde{A}(\omega)$ is uncorrelated QRPA propagator:
\be
\tilde{A}^{\vphantom{(-)}}_{12,34} (\omega) =
- \frac{
\eta^{\vphantom{(+)}}_1\,
\delta^{\vphantom{(+)}}_{\eta_{\mbts{1}},-\eta_{\mbts{2}}}\,
\delta^{\vphantom{(+)}}_{13}\,
\delta^{\vphantom{(+)}}_{24}}
{\omega - E_{12}}\,,
\qquad E_{12} = E_1 - E_2\,.
\label{atld}
\ee
To go beyond the QRPA we have to find reasonable approximations
for the quantities ${\cal U}^{\,e}$ and $\Sigma^{\,e}$.
In the present work we shall use a QRPA-based version of the
quasiparticle-phonon coupling (QPC) model (see Ref.~\cite{BM2}).
This model is discussed and used in a variety of papers:
see, e.g., Refs. \cite{S92,KTT,BBBD,CB01,IS74}.
Within the QPC model one can restrict oneself to so-called
$g^2$ approximation where $g$ is an amplitude of the
quasiparticle-phonon interaction
(see Ref.~\cite{KTT} for more details).
Under some simplifying assumptions,
this approximation can be obtained
in the GF method (see Refs.~\cite{E69,ES69,WE}).
Within the QPC model and $g^2$ approximation we have
the following formulas for the quantities
${\cal U}^{\,e}$ and $\Sigma^{\,e}$:
\bea
{\cal U}^{\,e}_{\,12,34} (\omega,\,\ve,\,\ve') &=&
\sum_{\eta, m}
\frac{\eta\,g^{m(\eta)^{\ds *}}_{31}\,g^{m(\eta)}_{42}}
{\ve - \ve' + \eta\,(\omega_m - i\!\cdot\!0)}\,,
\label{uephon}\\
\Sigma^{\,e}_{\,12}(\ve) &=&
\sum_{3,\eta, m} \frac{\delta_{\eta,\eta_{\mbts{3}}}
g^{m(\eta)^{\ds *}}_{13}\,g^{m(\eta)}_{23}}
{\ve - E_3 - \eta\,(\omega_m - i\!\cdot\!0)}\,,
\label{sgephon}
\eea
where $\eta = \pm 1\,$. Hereafter it is assumed that
the quasiparticle-phonon amplitudes $g^{m(\eta)}_{12}$
are related to the transition amplitudes $\rho^{m(\eta)}_{12}$
by means of QRPA equations:
\be
g^{m(\eta)}_{12} = \sum_{34}
\tilde{{\cal F}}^{\vphantom{A}}_{\vphantom{\bar{1}}12,34}\,
\rho^{m(\eta)}_{34}\,,
\qquad
\rho^{m(\eta)}_{12} =
\frac{\eta^{\vphantom{(+)}}_1\,
\delta^{\vphantom{(+)}}_{\eta_{\mbts{1}},-\eta_{\mbts{2}}}}
{\eta\,\omega_m - E_{12}}\,g^{m(\eta)}_{12}\,,
\label{qrpa1}
\ee
where $\tilde{{\cal F}}$ is an amplitude of the effective
interaction which generally differs from the amplitude
${\cal F}$ entering Eq.~(\ref{bseqren}).
Notice that the QRPA equations acquire very simple form in
the representation of single-quasiparticle $\psi$-functions.
The Eqs.~(\ref{qrpa1}) have to be supplemented by the
normalization condition
\be
\frac{1}{2} \sum_{12} \eta\,\eta^{\vphantom{(+)}}_1\,
\bigl| \rho^{m(\eta)}_{12} \bigr|^2 = 1
\label{qrpa2}
\ee
and by the condition of the antisymmetry
\be
\rho^{m(\eta)}_{\vphantom{\bar{1}}12} = -
\rho^{m(\eta)}_{\bar{2}\bar{1}}
\label{arho}
\ee
which is obviously fulfilled for the exact transition
amplitudes defined by Eq.~(\ref{tran}).
Notice, however, that in contrast to the case considered
in Refs. \cite{KTT,T89}, the quasiparticle-phonon amplitudes
$g^{m(\eta)}_{12}$ in Eqs. (\ref{uephon}) and (\ref{sgephon})
determine the coupling with excitations
both in the ph, and in the pp channels
[see Eqs.~(\ref{qrpa1}) and comments after Eq.~(\ref{tran})].

The QTBA in combination with Eqs. (\ref{uephon}) and
(\ref{sgephon}) leads to the following ansatz for the
correlated propagator:
\be
A^{\vphantom{(+)}}_{12,34} (\omega) = \sum_{5678} \bigl[\,
\delta^{\vphantom{(+)}}_{15}\,\delta^{\vphantom{(+)}}_{26} +
Q^{(+-)}_{12,56} (\omega)\,\bigr]\,
A^{(--)}_{56,78} (\omega)\,\bigl[\,
\delta^{\vphantom{(+)}}_{73}\,\delta^{\vphantom{(+)}}_{84} +
Q^{(-+)}_{78,34} (\omega)\,\bigr]\,
+ P^{(++)}_{12,34} (\omega)\,,
\label{dfprop}
\ee
where the upper indices denote products of the first and the
second pairs of lower $\eta$-indices. In particular, for
the component $A^{(--)}_{12,34}(\omega)$ of the propagator
we have: $\eta^{\vphantom{(+)}}_1 \eta^{\vphantom{(+)}}_2 =
\eta^{\vphantom{(+)}}_3 \eta^{\vphantom{(+)}}_4 = -1$.
This component is determined by the equation
\be
A^{(--)}_{12,34} (\omega) =
\tilde{A}^{\vphantom{(-)}}_{12,34} (\omega) -
\sum_{5678} \tilde{A}^{\vphantom{(-)}}_{12,56} (\omega)\,
\Phi^{\vphantom{(-)}}_{56,78} (\omega)\,
A^{(--)}_{78,34} (\omega)\,,
\label{ammeq}
\ee
where $\tilde{A}(\omega)$ is the QRPA propagator defined by
Eq.~(\ref{atld}). For the remaining quantities in the Eqs.
(\ref{dfprop}) and (\ref{ammeq}) we obtain
\bea
Q^{(+-)}_{12,34} (\omega) &=&
Q^{(+-)\,\mbsu{res}}_{12,34} (\omega) +
\delta^{\vphantom{(+)}}_{\eta_{\mbts{1}},\eta_{\mbts{2}}}\,
\delta^{\vphantom{(+)}}_{\eta_{\mbts{3}},-\eta_{\mbts{4}}}\,
\biggl(\,
\frac{\Sigma^{\,\mbsu{GSC}}_{\,31}}{E^{\vphantom{(+)}}_{31}}\,
\delta^{\vphantom{(+)}}_{24} -
\delta^{\vphantom{(+)}}_{31}\,
\frac{\Sigma^{\,\mbsu{GSC}}_{\,24}}{E^{\vphantom{(+)}}_{24}}
\biggr)\,,
\label{dfqpm}\\
Q^{(-+)}_{12,34} (\omega) &=&
Q^{(-+)\,\mbsu{res}}_{12,34} (\omega) -
\delta^{\vphantom{(+)}}_{\eta_{\mbts{1}},-\eta_{\mbts{2}}}\,
\delta^{\vphantom{(+)}}_{\eta_{\mbts{3}},\eta_{\mbts{4}}}\,
\biggl(\,
\frac{\Sigma^{\,\mbsu{GSC}}_{\,31}}{E^{\vphantom{(+)}}_{31}}\,
\delta^{\vphantom{(+)}}_{24} -
\delta^{\vphantom{(+)}}_{31}\,
\frac{\Sigma^{\,\mbsu{GSC}}_{\,24}}{E^{\vphantom{(+)}}_{24}}
\biggr)\,,
\label{dfqmp}\\
\vphantom{
\frac{\Sigma^{\,\mbsu{GSC}}_{\,24}}{E^{\vphantom{(+)}}_{24}}}
\Phi^{\vphantom{\mbsu{res}}}_{12,34} (\omega) &=&
\Phi^{\,\mbsu{res}}_{12,34} (\omega) +
\bar{\Phi}^{\,\mbsu{GSC}}_{12,34} +
\Phi^{\,\mbsu{GSC \it s.e.}}_{12,34} (\omega)\,.
\label{dfphi}
\eea
In these formulas,
a superscript ``res'' denotes the resonant parts
of the amplitudes, the quantities $\bar{\Phi}^{\,\mbsu{GSC}}$
and $\Phi^{\,\mbsu{GSC \it s.e.}} (\omega)$
represent contributions of the GSC. They consist of
the static part arising from the induced interaction
($\bar{\Phi}^{\,\mbsu{GSC}}$) and of the part arising from
the self-energy insertions ($\Phi^{\,\mbsu{GSC \it s.e.}}$):
\bea
\bar{\Phi}^{\,\mbsu{GSC}}_{12,34} &=& -
\delta^{\vphantom{(+)}}_{\eta_{\mbts{1}},-\eta_{\mbts{2}}}\,
\delta^{\vphantom{(+)}}_{\eta_{\mbts{3}},-\eta_{\mbts{4}}}\,
\sum_{\eta,m}\biggl(
\delta^{\vphantom{(+)}}_{\eta,\eta_{\mbts{3}}}\,
\rho^{m(\eta)}_{13}\,g^{m(\eta)^{\ds *}}_{24} +
\delta^{\vphantom{(+)}}_{\eta,\eta_{\mbts{4}}}\,
g^{m(\eta)}_{13}\,\rho^{m(\eta)^{\ds *}}_{24}\biggr)\,,
\label{phigsc1}\\
\Phi^{\,\mbsu{GSC \it s.e.}}_{12,34} (\omega) &=&
\eta^{\vphantom{(+)}}_1\,
\delta^{\vphantom{(+)}}_{\eta_{\mbts{1}},-\eta_{\mbts{2}}}\,
\delta^{\vphantom{(+)}}_{\eta_{\mbts{3}},-\eta_{\mbts{4}}}\,
\biggl(\,
\Sigma^{\,\mbsu{GSC}}_{\,31}\,
\bigl(\,\delta^{\vphantom{(+)}}_{24} +
q^{\vphantom{(+)}}_{24}\,\bigr) -
\bigl(\,\delta^{\vphantom{(+)}}_{31} +
q^{\vphantom{(+)}}_{31}\,\bigr)\,
\Sigma^{\,\mbsu{GSC}}_{\,24}
\nonumber\\
&& - \bigl(\,
q^{\vphantom{(+)}}_{31}\,\delta^{\vphantom{(+)}}_{24} +
\delta^{\vphantom{(+)}}_{31}\,q^{\vphantom{(+)}}_{24} +
q^{\vphantom{(+)}}_{31}\,q^{\vphantom{(+)}}_{24}\bigr)
\bigl[\,\omega - {\txts \frac{1}{2}}\,(E_{12} + E_{34})
\bigr] \biggr)\,,
\label{phigsc2}
\eea
where
\bea
\Sigma^{\,\mbsu{GSC}}_{\,12} &=& {\txts \frac{1}{2}}\,
\bigl(\,1 +
\delta^{\vphantom{(+)}}_{\eta_{\mbts{1}},-\eta_{\mbts{2}}}
\bigl)
\sum_{3,\eta,m} \eta\,
\delta^{\vphantom{(+)}}_{\eta,\eta_{\mbts{3}}}\,\biggl(
\rho^{m(\eta)^{\ds *}}_{13}g^{m(\eta)}_{23} +
g^{m(\eta)^{\ds *}}_{13}\rho^{m(\eta)}_{23}\biggr)\,,
\label{sgsc}\\
q^{\vphantom{(+)}}_{12} &=&
\sum_{3,\eta,m}
\delta^{\vphantom{(+)}}_{\eta,\eta_{\mbts{3}}}\,
\rho^{m(\eta)^{\ds *}}_{13}\rho^{m(\eta)}_{23}\,.
\label{q12}
\eea
The component $P^{(++)} (\omega)$ of the correlated propagator
and the resonant parts of the amplitudes entering Eqs.
(\ref{dfqpm})--(\ref{dfphi}) are defined as
\bea
P^{(++)}_{12,34} (\omega) &=&
\hphantom{-}
\sum_{5678,\eta,m}
\zeta^{\,m56(\eta)}_{\,12}\,
\tilde{A}^{(\eta)}_{56,78} (\omega-\eta\,\omega_m)\,
\zeta^{\,m78(\eta)^{\ds *}}_{\,34}\,,
\label{ppp}\\
Q^{(+-)\,\mbsu{res}}_{12,34} (\omega) &=&
\hphantom{-}
\sum_{5678,\eta,m}
\zeta^{\,m56(\eta)}_{\,12}\,
\tilde{A}^{(\eta)}_{56,78} (\omega-\eta\,\omega_m)\,
\gamma^{\,m78(\eta)^{\ds *}}_{\,34}\,,
\label{qpmres}\\
Q^{(-+)\,\mbsu{res}}_{12,34} (\omega) &=&
\hphantom{-}
\sum_{5678,\eta,m}
\gamma^{\,m56(\eta)}_{\,12}\,
\tilde{A}^{(\eta)}_{56,78} (\omega-\eta\,\omega_m)\,
\zeta^{\,m78(\eta)^{\ds *}}_{\,34}\,,
\label{qmpres}\\
\Phi^{\,\mbsu{res}}_{12,34} (\omega) &=&
-\sum_{5678,\eta,m}
\gamma^{\,m56(\eta)}_{\,12}\,
\tilde{A}^{(\eta)}_{56,78} (\omega-\eta\,\omega_m)\,
\gamma^{\,m78(\eta)^{\ds *}}_{\,34}\,,
\label{phires}
\eea
where
\bea
\gamma^{\,m56(\eta)}_{\,12} &=&
\delta^{\vphantom{(+)}}_{\eta,\eta_{\mbts{5}}}\,
\delta^{\vphantom{(+)}}_{\eta_{\mbts{1}},-\eta_{\mbts{2}}}\,
\delta^{\vphantom{(+)}}_{\eta_{\mbts{5}},-\eta_{\mbts{6}}}\,
\bigl(\,\delta^{\vphantom{(+)}}_{15}\,g^{m(\eta)}_{62}
-g^{m(\eta)}_{15}\delta^{\vphantom{(+)}}_{62}\,\bigr)\,,
\label{dfgamma}\\
\zeta^{\,m56(\eta)}_{\,12} &=&
\hphantom{\scs{-}}
\delta^{\vphantom{(+)}}_{\eta,\eta_{\mbts{5}}}\,
\delta^{\vphantom{(+)}}_{\eta_{\mbts{1}},\,\eta_{\mbts{2}}}\,
\delta^{\vphantom{(+)}}_{\eta_{\mbts{5}},-\eta_{\mbts{6}}}\,
\bigl(\,\delta^{\vphantom{(+)}}_{15}\,\rho^{m(\eta)}_{62}
-\rho^{m(\eta)}_{15}\delta^{\vphantom{(+)}}_{62}\,\bigr)\,,
\label{dfzeta}
\eea
$\tilde{A}^{(+)} (\omega)$ and $\tilde{A}^{(-)} (\omega)$
are the positive and the negative frequency parts of the
QRPA propagator defined by Eq.~(\ref{atld}), i.e.:
$\tilde{A} (\omega) =
\tilde{A}^{(+)} (\omega) + \tilde{A}^{(-)} (\omega)\,$,
\be
\tilde{A}^{(\eta)}_{12,34} (\omega) =
- \frac{\eta\,
\delta^{\vphantom{(+)}}_{\eta,\eta_{\mbts{1}}}\,
\delta^{\vphantom{(+)}}_{\eta_{\mbts{1}},-\eta_{\mbts{2}}}\,
\delta^{\vphantom{(+)}}_{13}\,
\delta^{\vphantom{(+)}}_{24}}
{\omega - \eta\,\eta^{\vphantom{(+)}}_1
E^{\vphantom{(+)}}_{12}}\,.
\label{atlde}
\ee

Correlated propagator $A(\omega)$ defined by Eq.~(\ref{dfprop})
and subsequent equations includes contributions of three types:
(i) pure 2q configurations associated with uncorrelated QRPA
propagator $\tilde{A}(\omega)$,
(ii) 2q$\otimes$phonon configurations introduced by
the quantities ${\cal U}^{\,e}$ and $\Sigma^{\,e}$, and
(iii) uncontrollable more complicated configurations arising
due to the GSC effects and their combinations with the
above-mentioned configurations.

\subsection{Sum rule analysis and a refinement of the model
            \label{sec33}}

The formulas of the previous subsection completely determine
the correlated propagator of the model
within the $g^2$ approximation.
By construction, this propagator
contains all the $g^2$ contributions, including those from GSC.
However, exact fulfillment of the sum rules in this approach
is not guaranteed. Let us consider this question in more detail.
Usually, the sum rule is meant in the sense of relation between
the moment $m_k$ of the strength function $S(E)$
and the ground state expectation value of certain operator
(see, e.g., Ref.~\cite{AL88}).
The moment $m_k$ is defined as
\be
m_k = \frac{1}{2}\,\int_{-\infty}^{\,\infty}S(E)\,E^k dE
\label{dfmk}
\ee
at $\Delta \to +0$ in Eq.~(\ref{dfstrf}).
Introducing asymptotic expansion of the exact response function:
\be
R_{12,34} (\omega) \sim -
\sum_{k=0}^{\infty} M^{\,(k)}_{12,34}\,\omega^{-k-1}
\label{asymp}
\ee
and using Eqs. (\ref{lrmre}), (\ref{spectl}),
(\ref{dfextv}), (\ref{dfstrf}),
one can show that the moments $m_k$ are expressed through
the coefficient functions $M^{\,(k)}_{12,34}$ by the formula:
\be
m_k = \frac{1}{4}\,\sum_{1234}\,
V^{\,0^{\ds *}}_{\,21}
M^{\,(k)}_{12,34}\;
V^{\,0^{\vphantom{^{\ds *}}}}_{\,43}\,.
\label{mkmk}
\ee
In particular, making use of the BSE (\ref{bseqr})
in the energy representation one can obtain:
\be
M^{\,(0)}_{12,34} =
\delta^{\vphantom{(0)}}_{31}\,
\rho^{\vphantom{(0)}}_{\,24} -
\rho^{\vphantom{(0)}}_{\,31}\,
\delta^{\vphantom{(0)}}_{24}\,,
\label{dfcf0}
\ee
where
$\rho^{\vphantom{(0)}}_{\,12} = \langle\, 0\, |\,
b^{\dag}_2\,b^{\vphantom{\dag}}_1\,|\, 0 \,\rangle\,$
is the extended density matrix (EDM).
Substituting Eq.~(\ref{dfcf0}) into the Eq.~(\ref{mkmk})
we get the so-called non-energy-weighted sum rule (NEWSR):
\be
m_0 = {\txts \frac{1}{4}}\,\mbox{Tr}
\bigl(\;\rho\,\bigl[\,V^{\,0},
V^{\,0^{\,\scs \dag}}\bigr]\,\bigr)\,.
\label{newsr}
\ee
Notice that the factor $\frac{1}{4}$ in this formula
(instead of the usual factor $\frac{1}{2}$) arises from
the definition (\ref{dfextv}) of the external field operator
in the extended space taken in the antisymmetric form.

Thus, in order to ensure exact fulfillment of the NEWSR
the coefficient function $M^{\,(0)}_{12,34}$
of the model must have the form (\ref{dfcf0}) with properly
normalized EDM $\rho$.
Using the formulas of the previous subsection and the
definition of the quantity $M^{\,(0)}_{12,34}$
through the expansion (\ref{asymp}) one can show that
Eq.~(\ref{dfcf0}) is not fulfilled in the model considered
and that the NEWSR is fulfilled only up to within the terms
of order $g^4$.

It is not difficult, however, to remedy this drawback
within the above-described scheme based on the
$g^2$ approximation. First of all, let us include
energy-independent operator $\Sigma^{\,\mbsu{GSC}}$ defined by
Eq.~(\ref{sgsc}) into the mean-field part $\tilde{\Sigma}$
of the total mass operator $\Sigma(\ve)$. It can be done
because the only constraint on the operator $\tilde{\Sigma}$
was the condition of its energy independence.
This redefinition of $\tilde{\Sigma}$ means that we have to use
in all the equations the operator
$\bar{\Sigma}^{\,e}(\ve) = \Sigma^{\,e}(\ve) -
\Sigma^{\,\mbsu{GSC}}$
instead of $\Sigma^{\,e}(\ve)$. The replacement of
$\Sigma^{\,e}(\ve)$ by $\bar{\Sigma}^{\,e}(\ve)$
leads to disappearance of all the
terms containing $\Sigma^{\,\mbsu{GSC}}$ in Eqs.
(\ref{dfqpm}), (\ref{dfqmp}), and (\ref{phigsc2}).
The remaining part of the amplitude
$\Phi^{\,\mbsu{GSC \it s.e.}}(\omega)$ in Eq.~(\ref{dfphi})
can be taken into account through the renormalization
of the QRPA propagator $\tilde{A}(\omega)$ within the
$g^2$ approximation.

To this aim let us introduce matrix $\tilde{Z}_{12,34}$
defined by the following equations:
\be
\sum_{56}
\tilde{Z}^{\vphantom{(+)}}_{12,56}\,
\tilde{Z}^{\vphantom{(+)}}_{56,34} =
\delta^{\vphantom{(+)}}_{31}\,\delta^{\vphantom{(+)}}_{24} -
q^{\vphantom{(+)}}_{31}\,\delta^{\vphantom{(+)}}_{24} -
\delta^{\vphantom{(+)}}_{31}\,q^{\vphantom{(+)}}_{24}\,,
\qquad
\tilde{Z}^{\vphantom{\ds *}}_{12,34} =
\tilde{Z}^{\,\ds *}_{\,34,12}\,,
\label{dfzt}
\ee
where the matrix $q^{\vphantom{(+)}}_{12}$ is defined by
Eq.~(\ref{q12}).
In addition it will be supposed that the matrix
$\tilde{Z}_{12,34}$ is positive-definite
that can be always fulfilled if all the eigenvalues $q_i$
of the matrix $q^{\vphantom{(+)}}_{12}$
satisfy condition: $q_i < \frac{1}{2}\,$
[notice that $q_i \geqslant 0$,
as follows from Eq.~(\ref{q12})].
Because according to Eqs. (\ref{qrpa1}) and (\ref{q12})
we have: $q^{\vphantom{(+)}}_{12}=O(g^2)\,$,
the pointed condition is consistent with the previous model
assumptions. Thus, from Eq.~(\ref{dfzt}) it follows that
\be
\tilde{Z}^{\vphantom{(+)}}_{12,34} =
\delta^{\vphantom{(+)}}_{31}\,\delta^{\vphantom{(+)}}_{24} -
{\txts \frac{1}{2}}\,\bigl(\,
q^{\vphantom{(+)}}_{31}\,\delta^{\vphantom{(+)}}_{24} +
\delta^{\vphantom{(+)}}_{31}\,q^{\vphantom{(+)}}_{24}\,
\bigr) + \,O(g^4)\,.
\label{ztprop}
\ee

Further, using Eqs. (\ref{atld}), (\ref{phigsc2})
(without terms containing operator $\Sigma^{\,\mbsu{GSC}}$),
and (\ref{ztprop}), we obtain:
\be
\tilde{A}(\omega) - \tilde{A}(\omega)\,
\Phi^{\,\mbsu{GSC \it s.e.}}(\omega)\,\tilde{A}(\omega) =
\tilde{Z}\;\tilde{A}(\omega)\,\tilde{Z} + \,O(g^4)\,.
\label{eqzaz}
\ee
It enables one to redefine correlated propagator replacing
Eq.~(\ref{dfprop}) by the following ansatz:
\be
A^{\vphantom{(+)}}_{12,34} (\omega) = \sum_{5678}
Z^{L}_{12,56} (\omega)\,
A^{(--)}_{56,78} (\omega)\,
Z^{R}_{78,34} (\omega)
+ P^{(++)}_{12,34} (\omega)\,,
\label{rdfprop}
\ee
where
\bea
Z^{L}_{12,34} (\omega) &=& \sum_{56} \bigl[\,
\delta^{\vphantom{(+)}}_{15}\,\delta^{\vphantom{(+)}}_{26} +
Q^{(+-)\,\mbsu{res}}_{12,56} (\omega)\,\bigr]\,
\tilde{Z}^{\vphantom{(+)}}_{56,34}\,,
\label{dfzl}\\
Z^{R}_{12,34} (\omega) &=& \sum_{56}
\tilde{Z}^{\vphantom{(+)}}_{12,56}\,\bigl[\,
\delta^{\vphantom{(+)}}_{53}\,\delta^{\vphantom{(+)}}_{64} +
Q^{(-+)\,\mbsu{res}}_{56,34} (\omega)\,\bigr]\,.
\label{dfzr}
\eea
In Eq.~(\ref{rdfprop}) the propagator $A^{(--)}(\omega)$
is determined by Eq.~(\ref{ammeq}) in which the amplitude
$\Phi(\omega)$ is now defined as
\be
\Phi^{\vphantom{\mbsu{res}}}_{12,34} (\omega) = \sum_{5678}
\tilde{Z}^{\vphantom{(+)}}_{12,56}\,\bigl[\,
\Phi^{\,\mbsu{res}}_{56,78} (\omega) +
\bar{\Phi}^{\,\mbsu{GSC}}_{56,78}\,\bigr]\,
\tilde{Z}^{\vphantom{(+)}}_{78,34}\,,
\label{rdfphi}
\ee
instead of the Eq.~(\ref{dfphi}).

It is easy to see that the propagator $A(\omega)$
defined by Eqs. (\ref{ammeq}), (\ref{dfzt}),
(\ref{rdfprop})--(\ref{rdfphi}) coincides
up to within the terms of order $g^4$ with the propagator
defined in the previous subsection
[see Eq.~(\ref{dfprop}) and subsequent equations].
On the other hand, assuming that the effective charge in
Eq.~(\ref{rrren}) is equal to unit operator and making use of
the expansion (\ref{asymp}), one can find that this new
redefined propagator leads to the following result for
the coefficient function $M^{\,(0)}_{12,34}$:
\be
M^{\,(0)}_{12,34} =
\eta^{\vphantom{(+)}}_1\,
\delta^{\vphantom{(+)}}_{\eta_{\mbts{1}},-\eta_{\mbts{2}}}\,
\bigl(\,
\delta^{\vphantom{(+)}}_{31}\,\delta^{\vphantom{(+)}}_{24} -
q^{\vphantom{(+)}}_{31}\,\delta^{\vphantom{(+)}}_{24} -
\delta^{\vphantom{(+)}}_{31}\,q^{\vphantom{(+)}}_{24}\,\bigr) +
\eta^{\vphantom{(+)}}_1\,
\delta^{\vphantom{(+)}}_{\eta_{\mbts{1}},\,\eta_{\mbts{2}}}\,
\bigl(\,\delta^{\vphantom{(+)}}_{31}\,q^{\vphantom{(+)}}_{24} -
q^{\vphantom{(+)}}_{31}\,\delta^{\vphantom{(+)}}_{24}\,\bigr)\,.
\label{cf0q}
\ee
Here the first term containing
$\delta^{\vphantom{(+)}}_{\eta_{\mbts{1}},-\eta_{\mbts{2}}}$
follows from Eq.~(\ref{dfzt}).
The second term arises from the contribution of the
component $P^{(++)}(\omega)$ in Eq.~(\ref{rdfprop}).
From the Eq.~(\ref{cf0q}) we obtain
that in the modified version of the
model the coefficient function $M^{\,(0)}_{12,34}$
has the form (\ref{dfcf0}) with correlated EDM
$\rho$ defined as
\be
\rho^{\vphantom{(+)}}_{12} =
\tilde{\rho}^{\vphantom{(+)}}_{12} +
\eta^{\vphantom{(+)}}_1\,q^{\vphantom{(+)}}_{12}\,,
\label{rhocor1}
\ee
where
$\tilde{\rho}^{\vphantom{(+)}}_{12} =
\delta^{\vphantom{(+)}}_{\eta_{\mbts{1}},-1}\,
\delta^{\vphantom{(+)}}_{12}$
is the EDM of the HFB theory in the representation of
$\psi$-functions. Therefore we conclude that
if the EDM (\ref{rhocor1}) is normalized by the usual condition
\be
\int dy'\,\delta^{\vphantom{(+)}}_{\chi',\,+1}\,
\delta^{\vphantom{(+)}}_{\tau',\,\tau} \sum_{12}
\psi^{\vphantom{*}}_1(y')\,\psi^*_2(y')\,
\rho^{\vphantom{(+)}}_{12} = N_{\tau}\,,
\label{rhonor}
\ee
the NEWSR is fulfilled exactly within the QTBA.
It is worthwhile mentioning that the EDM (\ref{rhocor1})
arising in the QTBA coincides with the correlated EDM
$\rho^{\,c}$ which can be obtained from the Dyson equation
(\ref{dyson1}) and the Eqs. (\ref{tgf1e}), (\ref{sgephon}),
(\ref{sgsc}) within $g^2$ approximation:
\be
\rho^{\vphantom{c}}_{\vphantom{\hat{i}}12} =
\rho^{\,c}_{\vphantom{\hat{i}}12} \equiv
\int_{-\infty}^{\,\infty} \frac{d \ve}{2 \pi i}\;
e^{i \ve \tau} \bigl[\,\tilde{G}(\ve) +
\tilde{G}(\ve)\,\bigl(\,\Sigma^{\,e}(\ve)
- \Sigma^{\,\mbsu{GSC}}\,\bigr)\,\tilde{G}(\ve)\,
\bigr]_{12}\,,\qquad
\tau \to +\,0\,.
\label{rhocor2}
\ee

Finally, note that the GSC effects included by means of
renormalization of the QRPA propagator within the QTBA with
the help of the matrix $\tilde{Z}$ are the same as
the effects included in the renormalized QRPA (RQRPA,
see, e.g., Ref.~\cite{KVC}). It is known (see Ref.~\cite{SKK})
that within the standard RQRPA, the Ikeda sum rule (being
a particular case of the NEWSR) is violated.
The above analysis allows us to understand the reason
of this violation. It follows from Eq.~(\ref{cf0q}) that
to satisfy the NEWSR within
the QTBA it is necessary to take into account contribution
of the term $P^{(++)}(\omega)$.
This term represents dynamic contributions of the GSC
into the QTBA propagator (\ref{rdfprop}), which cannot be
reduced to the renormalization of the QRPA propagator
$\tilde{A}(\omega)$ and which are absent in the RQRPA.

\subsection{Antisymmetrization of the equations\\
            and inclusion of the two-phonon configurations
            \label{sec34}}

As can be seen from the spectral expansion (\ref{spectl}),
the physical observables of the theory are completely
determined by the antisymmetric response function
$L(\omega)$. Within the QTBA this exact function is
approximated by the effective antisymmetric response function
$L^{\mbsu{eff}}(\omega)$ defined as
[cf. Eq.~(\ref{lrmre})]
\be
L^{\mbsu{eff}}_{12,34}(\omega) =
R^{\,\mbsu{eff}}_{12,34}(\omega) -
R^{\,\mbsu{eff}}_{\bar{2}\bar{1},34}(\omega)\,,
\label{leff}
\ee
where $R^{\mbsu{eff}}(\omega)$ is the solution of
Eq.~(\ref{bseqren}). It is easy to prove the following.
First, the polarizability $\Pi (\omega)$
[see Eqs.~(\ref{dfpol1})]
is actually determined by the function $L^{\mbsu{eff}}(\omega)$,
while symmetric part of the function $R^{\,\mbsu{eff}}(\omega)$
does not contribute to the Eqs.~(\ref{dfpol1}).
In other words, $\Pi (\omega)$ is invariant under the
transformation: $R^{\,\mbsu{eff}}(\omega) \to \frac{1}{2}\,
L^{\mbsu{eff}}(\omega)$.
Second, the function $L^{\mbsu{eff}}(\omega)$
is the doubled solution of the antisymmetrized QTBA equation
obtained from the Eq.~(\ref{bseqren}) with the help of
antisymmetrization of the correlated propagator $A(\omega)$.
This antisymmetrization can be implemented
by means of the following transformations
in Eqs. (\ref{ammeq}), (\ref{ppp})--(\ref{phires}):
$\tilde{A} (\omega) \to \frac{1}{2}\,\tilde{L}^0 (\omega)$,
$\tilde{A}^{(\eta)} (\omega) \to
\frac{1}{2}\,\tilde{L}^{0(\eta)} (\omega)$,
where $\tilde{L}^0 (\omega)$ is the antisymmetric
(uncorrelated) QRPA propagator, $\tilde{L}^{0(\eta)} (\omega)$
represents its positive and negative frequency parts
[cf. Eqs. (\ref{atld}) and (\ref{atlde})]:
\be
\tilde{L}^0_{12,34} (\omega) = \sum_{\eta = \pm 1}
\tilde{L}^{0(\eta)}_{12,34} (\omega)\,,\qquad
\tilde{L}^{0(\eta)}_{12,34} (\omega) =
\bigl(
\delta^{\vphantom{(+)}}_{\bar{2}3}\,
\delta^{\vphantom{(+)}}_{\bar{1}4} -
\delta^{\vphantom{(+)}}_{\vphantom{\bar{1}}13}\,
\delta^{\vphantom{(+)}}_{\vphantom{\bar{1}}24}
\bigr)\,
\frac{\eta\,
\delta^{\vphantom{(+)}}_{\eta,\eta_{\mbts{1}}}\,
\delta^{\vphantom{(+)}}_{\eta_{\mbts{1}},-\eta_{\mbts{2}}}}
{\omega - \eta\,\eta^{\vphantom{(+)}}_1
E^{\vphantom{(+)}}_{12}}\,.
\label{atldea}
\ee
We did not use the antisymmetric form of the QTBA equations
from the very beginning to simplify their derivation and
analysis. However, the antisymmetrization facilitates numerical
solution due to reduction of the dimensions of matrices
entering these equations.

The model described above
allows for the following straightforward extension related to
the definition of the resonant parts of the amplitudes entering
Eqs. (\ref{ammeq}), (\ref{rdfprop})--(\ref{rdfphi})
for the correlated propagator of the QTBA.
Contributions from these resonant parts
[defined by Eqs.(\ref{ppp})--(\ref{phires})]
to the response function describe simultaneous propagation of
the phonon and of the uncorrelated quasiparticle pair.
Natural generalization of this model is inclusion of the
correlations in the quasiparticle pair entering
2q$\otimes$phonon configuration, i.e. replacement of the
uncorrelated pair by the phonon. For the ph-channel BSE in
the normal Fermi system similar generalization, corresponding
to the replacement of the 1p1h$\otimes$phonon configurations
by the two-phonon intermediate states,
was discussed in Ref.~\cite{S76}. For the pp-channel BSE
analogous procedure has been implemented in Ref.~\cite{HEH77}.

Within the QTBA, two-quasiparticle correlations in
the 2q$\otimes$phonon intermediate states
(i.e., two-phonon configurations) can be included
in the following way.
Correlated counterpart of the above-defined quantity
$\tilde{L}^{0(\eta)} (\omega)$ is
$\tilde{L}^{(\eta)} (\omega)$ representing
the positive and the negative frequency parts of the
antisymmetric QRPA response function $\tilde{L} (\omega)$:
\be
\tilde{L}_{12,34} (\omega) = \sum_{\eta = \pm 1}
\tilde{L}^{(\eta)}_{12,34} (\omega)\,,\qquad
\tilde{L}^{(\eta)}_{12,34} (\omega) = - \sum_{n}
\frac{\eta\,\rho^{n(\eta)}_{12}\,\rho^{n(\eta)^{\ds *}}_{34}}
{\,\omega - \eta\,\omega_n}\,.
\label{reta}
\ee
In the Eq.~(\ref{reta}) it is supposed that the QRPA energies
$\omega_n$ and transition amplitudes $\rho^{n(\eta)}_{12}$
satisfy Eqs. (\ref{qrpa1})--(\ref{arho}).
The above considerations imply that transition
to the two-phonon configurations within the QTBA
can be accomplished by means of the replacement
$\tilde{A}^{(\eta)} (\omega) \to
\frac{1}{2}\,\tilde{L}^{(\eta)} (\omega)$
in Eqs. (\ref{ppp})--(\ref{phires}), which leads to the
following result:
\bea
P^{(++)}_{12,34} (\omega) &=&
- \frac{1}{2}
\sum_{\eta,m,n}\frac{\eta\,
\zeta^{\,mn(\eta)}_{\,12}\,
\zeta^{\,mn(\eta)^{\ds *}}_{\,34}}
{\omega-\eta\,\omega_{mn}}\,,
\label{pppc}\\
Q^{(+-)\,\mbsu{res}}_{12,34} (\omega) &=&
- \frac{1}{2}
\sum_{\eta,m,n}\frac{\eta\,
\zeta^{\,mn(\eta)}_{\,12}\,
\gamma^{\,mn(\eta)^{\ds *}}_{\,34}}
{\omega-\eta\,\omega_{mn}}\,,
\label{qpmresc}\\
Q^{(-+)\,\mbsu{res}}_{12,34} (\omega) &=&
- \frac{1}{2}
\sum_{\eta,m,n}\frac{\eta\,
\gamma^{\,mn(\eta)}_{\,12}\,
\zeta^{\,mn(\eta)^{\ds *}}_{\,34}}
{\omega-\eta\,\omega_{mn}}\,,
\label{qmpresc}\\
\Phi^{\,\mbsu{res}}_{12,34} (\omega) &=&
\hphantom{-} \frac{1}{2}
\sum_{\eta,m,n}\frac{\eta\,
\gamma^{\,mn(\eta)}_{\,12}\,
\gamma^{\,mn(\eta)^{\ds *}}_{\,34}}
{\omega-\eta\,\omega_{mn}}\,,
\label{phiresc}
\eea
where $\;\omega_{mn} = \omega_{m}+\omega_{n}\,$,
\be
\gamma^{\,mn(\eta)}_{\,12} = \sum_{56}
\gamma^{\,m56(\eta)}_{\,12}\,\rho^{\,n(\eta)}_{\,56}\,,
\qquad
\zeta^{\,mn(\eta)}_{\,12} = \sum_{56}
\zeta^{\,m56(\eta)}_{\,12}\,\rho^{\,n(\eta)}_{\,56}\,.
\label{dfgzc}
\ee

Physical arguments in favor of using Eqs.
(\ref{pppc})--(\ref{phiresc}) instead of
(\ref{ppp})--(\ref{phires}) are clear.
Notice, however, that the derivation
of the Eqs. (\ref{pppc})--(\ref{phiresc}) has not been
rigorous. It enables one only to assert that these formulas
recover the original Eqs. (\ref{ppp})--(\ref{phires}) in
the limit of vanishing quasiparticle interaction.
One can also show, using the completeness of the set of QRPA
transition amplitudes, that the NEWSR is fulfilled exactly
within two-phonon version of the QTBA.
The rigorous derivation of the Eqs.
(\ref{pppc})--(\ref{phiresc}) is based on the inclusion
of the additional (third order in the quasiparticle interaction)
contributions into the dynamic amplitude
${\cal U}^{\,e}$ defined by Eq.~(\ref{uephon}) and
will not be considered here.
It is worth noting that inclusion of the two-phonon
configurations in Eqs. (\ref{pppc})--(\ref{phiresc}) brings
the model closer to the QPM \cite{S92}.
Comparing the QTBA and the QPM one can infer
that treatment of the GSC within the QTBA is more consistent.
A more detailed comparison of these models is beyond the scope
of the present paper.

\subsection{Self-consistent scheme \label{sec35}}

Finally, we briefly outline the scheme which
enables one to eliminate spurious states within the QTBA.
These states, being a common problem of the microscopic
theories, are associated with existence of the non-trivial
external field operators $V^{\,0}$ satisfying the condition:
$[H,V^{\,0}]=[H^0,V^{\,0}]$, where $H$ and $H^0$ are the total
and the single-particle Hamiltonian, correspondingly
[see Eq.~(\ref{dfhtot})]. Elimination of the spurious states
within a consistent theory implies that they
must have zero excitation energy.
In terms of the GF method this means that the exact response
function $R(\omega)$ must have a pole at $\omega = 0$
corresponding to the spurious states. In particular,
it is well known that the QRPA response function
$\tilde{R}(\omega)$ satisfying the equation
\be
\tilde{R}(\omega) = \tilde{A}(\omega) -
\tilde{A}(\omega)\,\tilde{{\cal F}}\,
\tilde{R}(\omega)
\label{scs1}
\ee
has such a pole at $\omega = 0$ if the interaction amplitude
$\tilde{{\cal F}}$ is related to the mean-field operator
$\tilde{\Sigma}\,$ by the self-consistency condition
\be
\tilde{{\cal F}} =
\delta\,\tilde{\Sigma}\,/\,\delta\,\tilde{\rho}\,.
\label{scs2}
\ee

In the QTBA the situation is more complicated since the
correlated propagator $A(\omega)$ in Eq.~(\ref{bseqren})
has no simple structure of the QRPA propagator
$\tilde{A}(\omega)$. To avoid this difficulty
let us note that the exact solution of the Eq.~(\ref{bseqren})
with the propagator $A(\omega)$ defined by
Eqs. (\ref{rdfprop}), (\ref{ammeq})
can be represented in the form:
\bea
R^{\,\mbsu{eff}}(\omega)\!\!\! &=& \!\!\!
\bigl[\,1-P^{(++)}(\omega)\,{\cal F}^{\,P}(\omega)\,\bigr]\,
Z^L(\omega)\,\tilde{R}^{\,\mbsu{eff}}(\omega)\,Z^R(\omega)\,
\bigl[\,1-{\cal F}^{\,P}(\omega)\,P^{(++)}(\omega)\,\bigr]
\nonumber\\
&&+\;
P^{(++)}(\omega) -
P^{(++)}(\omega)\,{\cal F}^{\,P}(\omega)\,P^{(++)}(\omega)\,,
\label{scs3}
\eea
where energy-dependent interaction amplitude
${\cal F}^{\,P}(\omega)$ and the renormalized response function
$\tilde{R}^{\,\mbsu{eff}}(\omega)$ satisfy the equations:
\bea
{\cal F}^{\,P}(\omega) &=& {\cal F} -
{\cal F}\,P^{(++)}(\omega)\,{\cal F}^{\,P}(\omega)\,,
\label{scs4}\\
\tilde{R}^{\,\mbsu{eff}}(\omega) &=& \tilde{A}(\omega) -
\tilde{A}(\omega)\,\tilde{{\cal F}}(\omega)\,
\tilde{R}^{\,\mbsu{eff}}(\omega)\,,
\label{scs5}
\eea
with
\be
\tilde{{\cal F}}(\omega) =
Z^R(\omega)\,{\cal F}^{\,P}(\omega)\,Z^L(\omega) +
\Phi(\omega)\,.
\label{scs6}
\ee
If the amplitude $\tilde{{\cal F}}(0)$ coincides with the
interaction amplitude $\tilde{{\cal F}}$ satisfying
Eq.~(\ref{scs2}), the renormalized response function
$\tilde{R}^{\,\mbsu{eff}}(\omega)$
has the pole at $\omega = 0$ corresponding to the
spurious states by the same reasons as the QRPA
response function $\tilde{R}(\omega)$.
To ensure the fulfillment of the relationship
$\tilde{{\cal F}}(0) = \tilde{{\cal F}}$
we use the fact that the interaction amplitude ${\cal F}$
entering Eq.~(\ref{bseqren}) has not been constrained so far
by any conditions besides the property of its energy
independence. Let us now assume that the amplitude
${\cal F}$ satisfies the following equation:
\be
{\cal F} = {\cal F}^{\,P} +
{\cal F}^{\,P}P^{(++)}(0)\,{\cal F}\,,
\label{scs7}
\ee
where
\be
{\cal F}^{\,P} =
\bigl[Z^R(0)\bigr]^{-1}
\bigl[\,\tilde{{\cal F}} - \Phi(0)\,\bigr]\,
\bigl[Z^L(0)\bigr]^{-1},
\label{scs8}
\ee
and the amplitude $\tilde{{\cal F}}$ is determined by
Eq.~(\ref{scs2}).
If the Eqs. (\ref{scs7}) and (\ref{scs8}) are fulfilled,
then it follows from Eqs. (\ref{scs4}) and (\ref{scs6})
that $\tilde{{\cal F}}(0) = \tilde{{\cal F}}$.
Consequently, both function $\tilde{R}^{\,\mbsu{eff}}(\omega)$
and function $R^{\,\mbsu{eff}}(\omega)$
have the poles at $\omega = 0$ corresponding to the
spurious states. It means that these states are eliminated,
at least energetically,
within self-consistent version of the QTBA defined by the
above equations.

\section{CONCLUSIONS}

In this paper the problem of the microscopic description of
excited states of the even-even open-shell
atomic nuclei is considered.
The generalized Green function formalism (GGFF) has been
presented and used to formulate the model including
pairing, two-quasiparticle (2q), and more complex,
first of all quasiparticle-phonon correlations.
The GGFF is a modification
of the existing versions of Green function formalism which
is more suitable for solving the problem considered here.
Within the GGFF the normal and the anomalous Green functions
in the Fermi systems with pairing
are treated in a unified way in terms of the components of
generalized Green functions in a doubled space.
This treatment is analogous to the method used in
Ref.~\cite{IS77}. In the GGFF this method is extended to the
Fermi systems interacting through the
two-, three-, and other many-particle effective forces,
that is of importance for the nuclear physics where the
many-particle forces play an essential role.
Within the framework of this formalism the generalization
of the model of Ref.~\cite{T89} including the pairing
correlations has been developed.
The physical content of the model is determined by the
quasiparticle time blocking approximation (QTBA) which allows
one to keep the contributions of the 2q and 2q$\otimes$phonon
configurations, while excluding (blocking) more
complicated intermediate states.
It has been shown that within
the QTBA the non-energy-weighted sum rule is fulfilled exactly.
The model developed has been extended to include correlations in
the quasiparticle pair entering 2q$\otimes$phonon configuration,
i.e. to include two-phonon configurations.
Finally, the method of elimination of the spurious states
within the self-consistent QTBA has been considered.

\vspace{2em}
\noindent
This work was supported by the DFG (Germany) under grant
No. 436 RUS 113/806/0-1.
\vspace{2em}

%
%

%
\end{document}